\documentclass[12pt,a4paper]{article}
\usepackage{amsfonts}
\usepackage{amsmath}
\usepackage{bm}
\def \lam{\lambda}
\def \del{\partial}
\def \sig{\sigma}
\def \eps{\varepsilon}
\def \si{\sin \! }
\def \s2{\sin^2 \! }

\def \c2{\cos^2 \! }
\def \G{\Gamma}

\title{Several solutions of the Klein-Gordon equation in Kerr-Newman spacetime and the BSW effect}
\author{Hikaru Yumisaki\thanks{e-mail: yumisaki722@gmail.com}\\
\ \\
{\it Gunma University, 3-39-22 Showa, Maebashi 371-8511, Japan}}
\date{}
\begin{document}
\maketitle
\abstract{We investigate the radial part of the charged massive Klein-Gordon equation in Kerr-Newman spacetime, and in several specific situations, obtain exact solutions by means of essentially hypergeometric functions or their confluent types. Using these global solutions and generally obtained local solutions, we calculate a sort of intensity of the collision of two field excitations, which is a slight generalization of the trace of the stress tensor. We find that when the black hole is nonextremal, the intensity of the collision of two ingoing modes is bounded. However, in the extremal limit, more precisely $\hbar \kappa_H \rightarrow 0$, the upper bound grows so that when the frequency of one of the two modes satisfies the critical relation, the intensity of the collision at the horizon becomes unboundedly large. Furthermore, the intensity of the collision of ingoing and outgoing modes is always unbounded, as well as in the classical particle theory. Our results suggest that the BSW effect is inherited by the quantum theory. }
\tableofcontents

\section{Introduction}
In classical particle theory, it was shown in~\cite{BSW2009} that the collision energy of two particles in the center of mass frame can be unboundedly large at the event horizon of an extremal Kerr black hole on the equatorial plane if the energy $\eps $ and angular momentum $m$ of one of the particles satisfies the critical relation $\eps = \Omega_H m $, where $\Omega_H $ is the angular velocity of the black hole. This is called the BSW effect. The authors of~\cite{BSW2009} discussed whether it is possible for an extremal black hole to be used as an ultra-high energy particle collider and a probe of dark matter and the Planck-scale physics. Many other authors have revealed its universality. In~\cite{WLGF2010}, similar phenomena were observed in the Kerr-Newman spacetime for neutral particles, and in~\cite{HK2011}~\cite{LCJ2013} the authors investigated the BSW effect for nonequatorial orbits and found the limit to the latitude where the critical particles reach the event horizon. In~\cite{Zas2010b}, it was shown that nonrotating but extremely charged black holes also cause similar divergences of the collision energy for critically charged particles. Many authors have discussed the BSW effect in a variety of situations~\cite{LYLWL2011, ZJ2015, Zas2010a, TKAAS2013, SPF2014, ADAH2013, Hus2012, IHK2012, PJ2012, GSA2014, SA2011, YCLJ2012, SP2012, Frol2012, YLLWL2014, DLG2013, Saad2014, SH2013, PJ2014, Fern2014, Prad2015, HJM2015, GA2015, SB2015b, TAAS2015, SB2015, LCDJ2011, WLLC2010, Gala2013} and the possibilities of other processes of unboundedly high-energy collision,~\cite{GP2011, GP2015, GP2013}, and for a review see~\cite{HK2014}.  

In this paper, we consider the quantum effect of the BSW process. In a naive consideration, since the tunnel effect enables critical particles to reach the event horizon regardless of the potential barrier, the BSW effect may occur even in nonextremal black hole. On the other hand, quantum effects may modify the collision energy of two particles so that it has an upper bound. The main purpose of this paper is to discuss whether the quantum theory inherits the BSW effect or not. 

Since it is difficult to treat full quantum field theory even in the test field approximation, we work in the classical Klein-Gordon field theory in Kerr-Newman spacetime. Classical fields describe a part of quantum features of corresponding particles, e.g., the tunnel effect and uncertainty relations. 

The Klein-Gordon equation in Kerr-Newman spacetime can be separated in terms of each variable and both the radial and angular part arrive at solving confluent Heun equations for nonextremal black hole. Furthermore, for the extremal case, the radial part arrives at the double confluent Heun equation. Unfortunately, in general parameters, exact solutions of Heun-type equations that are represented in globally defined forms are unknown. On the other hand, local solutions of Heun's equation that are expressed as power series are obtained in the Frobenius method, as are those of the other Fuchsian ordinary differential equations~\cite{HDE}. In~\cite{BCV2014}~\cite{BMV2014}, the authors transformed both parts of the Klein-Gordon equation into canonical form of the (double) confluent Heun equation explicitly and the power series solutions were used for the survey of Hawking radiation. Their approximate global solutions have also been investigated. In~\cite{Unruh1976}~\cite{Det1980}, approximate solutions were obtained through what is called asymptotic matching method. In~\cite{RS1977}, solutions of a massive scalar field in the Kerr-Newman background are obtained by use of Whittaker functions, but they are valid only near the event horizon and infinity. In~\cite{Hod2014b}, the author obtained the solutions in near-extremal Kerr-Newman spacetime for an electrically charged massive Klein-Gordon field that has critical frequency using the (confluent) hypergeometric functions. When it comes to the exact solutions, in~\cite{WC1999} the authors pointed out that the Klein-Gordon equation relates to various well-known equations in several specific cases. In particular, they contain the situations in which solutions are obtained by use of the Jacobi functions or the Legendre functions in the nonextremal case. In this paper, we search for several specific parameters in which the radial part of the charged massive Klein-Gordon equation in Kerr-Newman spacetime reduces to essentially hypergeometric equations or their confluent types. As a result, we obtain the globally defined exact solutions in each of the cases. Our results contain those of~\cite{Hod2014b} and~\cite{WC1999}. Related works in different approaches are found in~\cite{Kra2016}. 

For our purpose, as long as the qualitative features of the BSW effect are concerned, the local behaviors of the solutions in the vicinity of the horizon are of importance rather than the global ones. However, the global solutions enable quantitative discussions.

This paper is organized as follows. In section \ref{revBSW}, we give an overview of the BSW effect in a somewhat general setting, namely, in Kerr-Newman spacetime for charged massive particles. In section \ref{field}, we see the transformations of the Klein-Gordon equation to the confluent and double confluent Heun equation, and obtain local solutions explicitly. In section \ref{Reduction}, we find several specific situations in which the radial part of the Klein-Gordon equation reduces to essentially hypergeometric equations or their confluent types. We obtain appropriately normalized solutions in each of the cases. In section \ref{trace}, we define a slight generalization of the trace of the stress tensor that can be used as a kind of intensity of ``collision'' of two field excitations, and evaluate it. In section \ref{conclusion}, we discuss what our results mean.


\section{An overview of the BSW effect}\label{revBSW}
In this section, we review some properties of geometry and orbits of charged massive particles in Kerr-Newman spacetime, and the BSW effect. 

\subsection{The Kerr-Newman geometry}
Kerr-Newman spacetime is generated by a black hole with mass $M$, angular momentum per unit mass $a = J / M$, and electric charge $Q$. The line element $d s$ and gauge field $\bm{A}$ in Kerr-Newman spacetime in Boyer-Lindquist coordinates are given by~\cite{BL1967}~\cite{Gravitation}
\begin{eqnarray}
ds^2 \!&=&\! - \frac{\Delta}{\rho^2 } \left( dt - a \s2\theta d\phi \right)^2 + \frac{\rho^2 }{\Delta} dr^2 + \rho^2 d\theta^2 + \frac{\s2\theta}{\rho^2 } \left[ (r^2 + a^2 )d\phi - a dt \right]^2 , \label{lineelement} \\
\bm{A} \!&=&\! A_a dx^a = - \frac{Q r}{\rho^2 } \left( dt - a \s2\theta d\phi \right) , \label{gaugefield}
\end{eqnarray}
where 
\begin{eqnarray}
\rho^2 = r^2 + a^2 \cos^2 \!\theta , 
\ \ \ \ \ \  \Delta = r^2 - 2Mr + a^2 +Q^2 .
\end{eqnarray}

If $M^2 > a^2 + Q^2 $, then $\Delta $ vanishes at $r = r_H = M + \sqrt{M^2 -a^2 -Q^2 }$ and $r = r_C = M - \sqrt{M^2 -a^2 -Q^2 }$, where $r = r_H$ and $r = r_C $ correspond to the event horizon and the Cauchy horizon, respectively. The geometry has a ring singularity at $(r,\ \theta ) = (0, \ \frac{\pi}{2})$. 

If $M^2 = a^2 + Q^2 $, then $\Delta $ has a double root $r = r_H = r_C $. In this case, it is said that the black hole is `extremely rotating and charged', or simply ``extremal''. 

If $M^2 < a^2 + Q^2 $, then the geometry has a ``naked singularity''. In this paper, we do not consider this situation. 

The angular velocity $\Omega_H $, electric potential $\Phi_H $ and surface gravity $\kappa_H $ at the event horizon are given by 
\begin{eqnarray}
\Omega_H \!&=&\! \frac{a}{r_H ^2 + a^2 } , \\
\Phi_H \!&=&\! \frac{Q r_H }{r_H ^2 + a^2 } , \\
\kappa_H \!&=&\! \frac{r_H - r_C }{2(r_H ^2 + a^2 )} . 
\end{eqnarray}

\subsection{The motion of a charged massive particle in Kerr-Newman spacetime}
The motion of a charged particle is given by the action functional written in the form 
\begin{eqnarray}
S[x^a , \eta ] = \int d\lam \, \frac{1}{2} \Big[ \eta^{-1} g_{ab} \Big( \frac{dx^{a}}{d \lam} \Big) \Big( \frac{dx^{b}}{d \lam} \Big) - \eta \mu^2 + eA_a \frac{dx^a}{d\lam } \Big] , 
\end{eqnarray} 
where $\mu$ is the rest mass, $e$ is the electric charge, $\lam$ is a parameter of the orbit, $\eta$ is an auxiliary degree of freedom called the ``einbein''. It is clear that the motions of massless particles are obtained by taking the limit $\mu \rightarrow 0$. 

This system is singular in that the Lagrangian does not contain the derivative of $\eta$, i.e., some constraint conditions should be imposed to describe the system in the Hamiltonian formulation. The primary and secondary constraints are 
\begin{eqnarray}
&\pi_{\eta} \approx 0 & \\
&\{ \pi_{\eta} , H \} = - \frac{1}{2} [g^{ab} (\pi_a - eA_a )( \pi_b - eA_b ) + \mu^2 ] \approx 0 , & \label{secondconst}
\end{eqnarray}
respectively. Here, $\pi_q $ means the canonical momentum conjugate to a canonical variable $q$. We fix $\eta = 1$ so that $\lam$ is an affine parameter, $ds = \mu\, d\lam$, $p_a = \mu g_{ab} u^b = \pi_a - eA_a $, and $u^a = \mu^{-1} \frac{dx^a}{d\lam}$, where $\pi_a $ means the canonical momentum conjugate to $x^a $. 

The secondary constraint (\ref{secondconst}) gives the Hamilton-Jacobi equation of a particle with mass $\mu$ and charge $e$ in Kerr-Newman spacetime: 
\begin{eqnarray}
&&\frac{1}{\rho^2 }\bigg\{ -\frac{1}{\Delta }\bigg[ (r^2 + a^2 )\frac{\del S^{\rm HJ}}{\del t} + a \frac{\del S^{\rm HJ}}{\del \phi } + e Q r \bigg]^2 + \Delta \bigg( \frac{\del S^{\rm HJ}}{\del r} \bigg)^2 \nonumber \\
&& \ \ \ \ \ \ \ \ \ \ \ \ + \frac{1}{\sin^2 \! \theta }\bigg[ \frac{\del S^{\rm HJ}}{\del \phi } + a\sin^2 \! \theta \frac{\del S^{\rm HJ}}{\del t}\bigg]^2 + \bigg( \frac{\del S^{\rm HJ}}{\del \theta } \bigg)^2 \bigg\} + \mu^2 = 0 , \label{HJ}
\end{eqnarray}
where the Hamilton-Jacobi function is denoted by $S^{\rm HJ}$. $S^{\rm HJ}$, and is related to the canonical momenta $\pi_a$ conjugate to $x^a $ as $\pi_a = \del_a S^{\rm HJ}$. Since $t$ and $\phi$ are cyclic coordinates, Eq.\,(\ref{HJ}) is separable on the following assumption: 
\begin{eqnarray}
S^{\rm HJ} (t, \phi , r, \theta ) = - \eps t + m\phi + S^{\rm HJ}_r (r) + S^{\rm HJ}_ \theta (\theta ) , \label{sep}
\end{eqnarray}
where $\eps$ and $m$ are constants that correspond to the Killing energy and angular momentum, respectively. Substituting Eq.\,(\ref{sep}), Eq.\,(\ref{HJ}) is separated into left and right side in terms of variables $r$ and $\theta$. Both sides are equal to a separation constant $\mathcal{K}$: 
\begin{eqnarray}
&&-\Delta \bigg( \frac{dS^{\rm HJ}_r }{dr} \bigg)^2 - \mu ^2 r^2 + \frac{\Big[ (r^2 + a^2 )\eps  - am - e Q r \Big]^2 }{\Delta } \\
&\ & \ \ \ \ \ \ = \bigg( \frac{dS^{\rm HJ}_\theta }{d\theta } \bigg)^2 + \mu ^2 a^2 \cos ^2 \! \theta + \frac{1}{\sin^2 \! \theta } ( m-a\eps \s2\theta )^2 \label{Ktheta} \\
&\ & \ \ \ \ \ \ = \mathcal{K} \ge 0 . 
\end{eqnarray}
Note that $\mathcal{Q} = \mathcal{K} - (m - a \eps )^2 $ is called Carter's constant. The Hamilton-Jacobi function is integrated as 
\begin{eqnarray}
S^{\rm HJ}_r = \sig_r \int^r dr \frac{\sqrt{R(r)}}{\Delta }, \ \ \ \ \ \  S^{\rm HJ}_\theta = \sig_\theta \int^\theta d\theta \sqrt{\Theta (\theta )} , 
\end{eqnarray}
where the choices of the two signs $\sig_r = \pm 1$ and $\sig_{\theta} \pm 1$ are independent and 
\begin{eqnarray}
R(r) \!&=&\! P(r)^2 - \Delta ( \mu ^2 r^2 + \mathcal{K} ) , \\
\Theta (\theta ) \!&=&\! \mathcal{K} - \mu ^2 a^2 \cos^2 \!\theta - B(\theta )^2 , \\
P(r) \!&=&\! (r^2 +a^2 )\eps  -am - e Q r , \label{defP}\\
B(\theta ) \!&=&\! \frac{1}{\sin \theta} (m - a \eps \sin^2 \! \theta ) . \label{defB}
\end{eqnarray}
The classical motions are permitted only when
\begin{eqnarray}
R(r) \ge 0 , \label{cond1}\\
\Theta (\theta) \ge 0 \label{cond2}
\end{eqnarray}
are satisfied. 

We can write $\pi_a = \del S^{\rm HJ} / \del x^a$ in the following form: 
\begin{eqnarray}
\pi_t \!&=&\! - \eps , \label{pt} \\
\pi_{\phi} \!&=&\! m , \label{pph} \\
\pi_r \!&=&\! \sig_r \frac{\sqrt{R}}{\Delta} , \label{pr} \\
\pi_{\theta} \!&=&\! \sig_{\theta} \sqrt{\Theta} , \label{pth}
\end{eqnarray}
and obtain $dx^a / d\lam = g^{ab} (\pi_b - eA_b ) $: 
\begin{eqnarray}
\rho^2 \frac{dt}{d\lam } \!&=&\! a B \sin\theta + \frac{(r^2 + a^2 )P}{\Delta } , \label{ut} \\
\rho^2 \frac{dr}{d\lam } \!&=&\! \sig_r \sqrt{R} , \\
\rho^2 \frac{d\theta }{d\lam } \!&=&\! \sig_\theta \sqrt{\Theta } , \\
\rho^2 \frac{d\phi }{d\lam } \!&=&\! \frac{B}{\sin \theta} + \frac{aP}{\Delta } .
\end{eqnarray}
Nevertheless we have fixed $\eta = 1$, the orientation of the parameterization remains to be decided. For example, two orbits whose 4-momenta are $p_a$ and $- p_a$ are equivalent, for the difference between them is merely the orientation of the parameterization $\lam$. Therefore, one should identify $p_a $ with $- p_a $. To eliminate this ambiguity, we shall employ the future-directed condition $\frac{dx}{d\lam} \ge 0$, where $dx$ is a future-directed timelike 1-form. 

From the line element (\ref{lineelement}), one finds that the basis $(dt - a\sin^2 \theta d\phi , dr , d\theta , (r^2 + a^2 ) d\phi - a dt )$ diagonalizes the metric tensor $g_{ab} $, and $dt - a\sin^2 \theta d\phi $ is a future-directed timelike 1-form outside the event horizon. Thus, the future-directed condition is 
\begin{eqnarray}
\frac{dt - a \sin^2 \theta d\phi }{d\lam } = \frac{P}{\Delta } \ge 0 , \label{FDcond} 
\end{eqnarray}
i.e., $P(r) \ge 0$ outside the event horizon. 

\subsubsection{Permitted motions near the horizon}\label{Pmh} 
In all the sections of this paper, for simplicity, we use the symbols $P_H $, $P^{\prime }_H $, and $P^{\prime \prime }_H $ to denote the quantities 
\begin{eqnarray}
P_H \!\!&=&\!\! P(r_H) = (r_H ^2 + a^2 ) \eps - am - eQ r_H , \\
P^{\prime }_H \!\!&=&\!\! P^{\prime } (r_H ) = 2 r_H \eps - eQ , \\
P^{\prime \prime }_H \!\!&=&\!\! P^{\prime \prime }(r_H ) = 2 \eps ,  
\end{eqnarray}
and $P_C $, $P^{\prime }_C $ replacing $r_H $ by $r_C $. 

A particle thrown from outside the event horizon, say $r = r_0$, reaches the event horizon if and only if the four constants of the motion $\mu$, $\eps$, $m$, and $\mathcal{K}$ satisfy the existence conditions (\ref{cond1}) and (\ref{cond2}) in the entire region $r_H \le r \le r_0 $. 

Near the event horizon, 
\begin{eqnarray}
R(r) \!&\simeq&\! P_H  ^2 + 2 \big[ (P^{\prime }_H P_H - (r_H -M) (\mu^2 r_H ^2 + \mathcal{K}) \big] (r - r_H ) + \mathcal{O}\big( (r-r_H )^2 \big) . 
\end{eqnarray}
Thus, in the near-horizon limit $r \rightarrow r_H $, 
\begin{eqnarray}
R(r) \!&\rightarrow&\! R(r_H ) = P_H ^2 \ge 0, \\
R'(r) \!&\rightarrow&\! R^{\prime } (r_H ) = 2 P^{\prime }_H P_H - 2 (r_H -M) (\mu^2 r_H ^2 + \mathcal{K}). 
\end{eqnarray}

Particles that satisfy $P_H  = 0$ are called  ``critical'' particles. 
\begin{enumerate}
\item{
If a particle is noncritical, $R(r_H) > 0$, so there always exist regions from which particles reach the event horizon as long as $\Theta (\theta ) \ge 0$. 
	}
\item{
If a particle is critical, and the Kerr-Newman geometry is nonextremal, $R(r_H) = 0$ and $R'(r_H) < 0$, so there are no such regions: the critical particles never reach the event horizon. 

However, the situation is somewhat subtle. For a slightly noncritical particle, i.e., $P_H ^2 \ne 0$ but sufficiently small, the region from which the particle reaches the event horizon is 
\begin{eqnarray}
r_H \le r \le r_0 \simeq \frac{P_H ^2 }{2(r_H - M)(\mu^2 r_H ^2 + \mathcal{K})} + r_H . 
\end{eqnarray}
	}
\item{
If the particle is critical, and the Kerr-Newman geometry is extremal, $R(r_H) = R'(r_H) = 0$ and 
\begin{eqnarray}
R^{\prime \prime }(r_H ) = 2 \big[ (P^{\prime }_H )^2 - (\mu^2 M^2 + \mathcal{K} ) \big] , 
\end{eqnarray}
so the condition for the critical particle to reach the horizon is 
\begin{eqnarray}
\mathcal{K} \le (P^{\prime }_H )^2 - \mu^2 M^2 \label{conditionK}
\end{eqnarray}
and from condition (\ref{cond2}), 
\begin{eqnarray}
\mathcal{K} \ge \mu^2 a^2 \cos^2 \!\theta + B(\theta )^2 . 
\end{eqnarray}
Thus, the condition for such a $\mathcal{K}$ to exist is 
\begin{eqnarray}
\mu^2 a^2 \cos^2 \!\theta + B(\theta )^2 \le (P^{\prime }_H )^2 - \mu^2 M^2 . \label{belt}
\end{eqnarray}
This condition (\ref{belt}) for $e = Q = 0$ was investigated in~\cite{HK2011}, and for $e = 0$, $Q \ne 0$ in~\cite{LCJ2013}, and gives $\s2\theta $ a lower bound (the region is called the high-velocity collision belt) in each case. 
	}
\end{enumerate}


\subsection{The collision energy of two particles}
Consider two particles whose 4-velocities are $u_i^a$, where $i = 1, 2$. We distinguish the particles by subscript index $i = 1, 2$. The collision energy $E_{\rm cm}$ of the two particles in the center-of-mass frame is defined by~\cite{BSW2009}
\begin{eqnarray}
E_{\rm cm}^2 \!&=&\! -g_{ab}(\mu_1 u_1^a + \mu_2 u_2^a)(\mu_1 u_1^b + \mu_2 u_2^b) \\
\!&=&\! \mu_1 ^2 + \mu_2 ^2 - 2 g^{ab } (\pi_{1a} - e A_a )(\pi_{2b} - e A_b ) . \label{defEcm}
\end{eqnarray}
Substituting Eqs.\,(\ref{gaugefield}), (\ref{pt}), (\ref{pph}), (\ref{pr}), and (\ref{pth}), one obtains 
\begin{eqnarray}
E_{\rm cm} ^2 \!&=&\! \mu _1^2 + \mu _2^2 + \frac{2}{\rho^2 }\bigg[ \frac{P_1 P_2 - \sig_{r1}\sig_{r2}\sqrt{R_1}\sqrt{R_2}}{\Delta } - B_1 B_2 - \sig_{\theta 1}\sig_{\theta 2}\sqrt{\Theta_1 }\sqrt{\Theta_2 } \bigg] . \label{Ecm}
\end{eqnarray}
One can see that $E_{\rm cm}$ can diverge when $\Delta = 0$ or $\rho^2 = 0$. 

\subsubsection{Near-horizon limit}
Let us take the near-horizon limit $r \rightarrow r_H$. For simplicity, we write $P_{Hi} = P_i (r_H ) = (r_H ^2 + a^2 )\eps_i - a m_i - e_i Q r_H $ $(i = 1,\, 2)$ and $\rho_H ^2 = \rho^2 (r_H ) = r_H ^2 + a^2 \cos^2 \!\theta$.  

From Eq.\,(\ref{Ecm}), naively it seems that $E_{\rm cm} ^2 $ diverges in the limit $\Delta \rightarrow 0$. However, it is an indeterminate form. The asymptotic behaviors of $P_1 P_2 / \Delta $ and $\sqrt{R_1 } \sqrt{R_2 } / \Delta $ are 
\begin{eqnarray}
\frac{P_1 P_2 }{\Delta } \!\!\!\!&\simeq &\!\!\!\! \frac{P_{H1} P_{H2} }{(r_H - r_C )(r - r_H )} - \frac{P_{H1} P_{H2} }{(r_H - r_C )^2 } + \frac{P^{\prime }_{H1} P_{H2} + P_{H1} P^{\prime }_{H2} }{r_H - r_C } + \mathcal{O}(r - r_H ) , \\
\frac{\sqrt{R_1 } \sqrt{R_2 }}{\Delta } \!\!\!\!&\simeq &\!\!\!\! \frac{P_{H1} P_{H2} }{(r_H - r_C )(r - r_H )} - \frac{P_{H1} P_{H2} }{(r_H - r_C )^2 } + \frac{P^{\prime }_{H1} P_{H2} + P_{H1} P^{\prime }_{H2} }{r_H - r_C } \nonumber \\
	&& - \frac{1}{2} (\mu_1 ^2 r_H ^2 + \mathcal{K}_1 ) \frac{P_{H2}}{P_{H1}} - \frac{1}{2} (\mu_2 ^2 r_H ^2 + \mathcal{K}_2) \frac{P_{H1}}{P_{H2}} + \mathcal{O}(r - r_H ) . 
\end{eqnarray}

\ \\
i)When $\sig_{r1} \sig_{r2} = 1$, the two particles are both ingoing or both outgoing. In the near-horizon limit $\Delta \rightarrow 0$, unless $P_{H1} = 0$ or $P_{H2} = 0$, the collision energy converges: 
\begin{eqnarray}
\lim_{r \rightarrow r_H } E_{\rm cm}^2 \!&=&\! \mu _1^2 + \mu _2^2 + \frac{1}{\rho_H ^2 }\bigg[ (\mu _1^2 r_H^2 + \mathcal{K}_1)\frac{P_{H2}}{P_{H1}} + (\mu _2^2 r_H^2 + \mathcal{K}_2)\frac{P_{H1}}{P_{H2}} \nonumber \\
&& -2B_1 B_2 - 2\sig_{\theta 1 }\sig_{\theta 2 }\sqrt{\Theta_1 }\sqrt{\Theta_2 }  \bigg] . \label{EcmNH1}
\end{eqnarray}
When $P_{H1}=0$ or $P_{H2}=0$, the collision energy at the horizon can diverge. As we defined in section \ref{Pmh}, particles that satisfy $P_{H} = (r_H ^2 + a^2 )\eps - a m - e Q r_H =0$ are called critical particles. 

As we saw in section \ref{Pmh}, critical particles reach the event horizon from an outer region only in the extremal case. In nonextremal case, slightly noncritical particles reach the event horizon from the region 
\begin{eqnarray}
r_H \le r \le r_0 \simeq \frac{P_H ^2 }{2(r_H - M )(\mu^2 r_H ^2 + \mathcal{K})} + r_H . \label{inner region}
\end{eqnarray}
Therefore, even in nonextremal case, the collision energy can be arbitrarily large, but the larger it is, the narrower the region in which the orbit can exist is. In~\cite{GP2011}, the authors pointed out that the collision energy of a ``multiple scattering process'' has no upper bound in which the first ordinary particle collides with the second particle in the region (\ref{inner region}) and gets energy and momentum so that the new $P_H $ of the second particle becomes closer to $0$ than the original one, and the second collision with another ordinary particle occurs on the event horizon.  

\ \\
ii)When $\sig_{r1} \sig_{r2} = -1$, the particles are ingoing and outgoing. In the near-horizon limit, unless both particles are critical, the collision energy diverges as 
\begin{eqnarray}
E_{\rm cm} ^2 \!&\simeq &\! \mu _1^2 + \mu _2^2 + \frac{1}{\rho_H ^2 } \bigg[ \frac{4 P_{H1} P_{H2}  }{(r_H - r_C)(r - r_H)}  \nonumber \\ 
&& - \frac{4P_{H1} P_{H2}}{(r_H - r_C)^2 } + \frac{4P^{\prime }_{H1} P_{H2} + 4P_{H1} P^{\prime }_{H2} }{r_H - r_C }  \nonumber \\ 
&& - (\mu _1^2 r_H^2 + \mathcal{K}_1)\frac{P_{H2}}{P_{H1}} - (\mu _2^2 r_H^2 + \mathcal{K}_2)\frac{P_{H1}}{P_{H2}}  \nonumber \\ 
&& -2B_1 B_2 - 2\sig_{\theta 1 }\sig_{\theta 2 }\sqrt{\Theta_1 }\sqrt{\Theta_2 }  \bigg] + \mathcal{O}(r - r_H ) \label{EcmNH2b} \\
\!&\approx&\! \frac{4 P_{H1} P_{H2}  }{\rho_H ^2 \Delta } + \mathcal{O}(\Delta^0). \label{EcmNH2}
\end{eqnarray}

\ \\
We itemize a summary of this section:  
\begin{enumerate}
	\item In classical particle theory, Eq.\,(\ref{EcmNH1}) tells us that in both extremal and nonextremal Kerr-Newman spacetime, the collision energy of two ingoing particles can be unboundedly large at the event horizon if one of the particles is critical, i.e., it satisfies $P_H = 0$. 
	\item However, critical particles are able to reach from outside to the event horizon only in the extremal case. 
	\item Equation (\ref{EcmNH2b}) tells us that the collision energy of an ingoing and an outgoing particle is unboundedly large on the horizon unless both particles are critical. 
\end{enumerate}

\ \\

In the following sections, we see the field-theoretical counterparts of above results. To begin with, let us investigate the field equation of a scalar field, namely, the Klein-Gordon equation in Kerr-Newman spacetime, and its solutions. 


\section{Charged massive scalar field in Kerr-Newman spacetime}\label{field}
\subsection{The Klein-Gordon and Hamilton-Jacobi equation}
The Klein-Gordon equation\footnote{This is minimally coupled with scalar curvature $R$. Since that of Kerr-Newman spacetime is $0$, the method of the coupling does not matter.} for a massive and electrically charged scalar field $\Psi$ in curved spacetime is written as 
\begin{eqnarray}
\bigg[ \frac{1}{\sqrt{-g}}D_a (g^{ab} \sqrt{-g} D_b) - \frac{\mu ^2 }{\hbar^2 }\bigg] \Psi = 0 , \label{KG}
\end{eqnarray}
where $D_a = \del_a - \displaystyle \frac{i e}{\hbar } A_a $, $e$ and $\mu$ are the electric charge and mass of the scalar field $\Psi$, respectively, and $g$ is the determinant of the metric tensor. For Kerr-Newman spacetime, 
\begin{eqnarray}
g = - \rho^4 \sin^2 \theta . 
\end{eqnarray}

To see the relation to the classical theory, we write $\Psi$ in the following form: 
\begin{eqnarray}
\Psi = \exp\Big( \frac{i}{\hbar } S \Big) . \label{Psi}
\end{eqnarray}
Substituting Eq.\,(\ref{Psi}) into Eq.\,(\ref{KG}), the equation can be arranged as 
\begin{eqnarray}
g^{ab} ( \del_a S - e A_a ) ( \del_b S - e A_b ) + \mu^2 = \frac{i \hbar }{\sqrt{- g}} \del_a \left[ g^{ab} \sqrt{- g} (\del_b S - e A_b ) \right] \label{KGS} . 
\end{eqnarray}
In the classical limit $\hbar \rightarrow 0$, $S$ formally satisfies the Hamilton-Jacobi equation 
\begin{eqnarray}
g^{ab} ( \del_a S - e A_a ) ( \del_b S - e A_b ) + \mu^2 = 0 .
\end{eqnarray}
Therefore, $S$ can be thought of as a candidate of a field-theoretical counterpart of the Hamilton-Jacobi function $S^{\rm HJ}$. 

Note that while the 4-gradient of the Hamilton-Jacobi function $\del_a S^{\rm HJ}$ represents the canonical momentum of the particle, that of the phase of the scalar field $\del_a S$ is not always the counterpart. Since the Klein-Gordon equation is a second-order differential equation, the general solution has two arbitrary constants, which relate to the arbitrariness of the superposition and normalization. That is why in general a configuration of the scalar field corresponds to a situation in which there are more than one particles having momenta different from each other. Therefore, to see whether a solution $\Psi $ corresponds not to a superposition of many momenta but to a single momentum at a point $x = x_0$, one should compare the asymptotic behaviors of $\del_a S$ and $\del_a S^{\rm HJ}$ around $x = x_0 $, calculating the logarithmic derivative of $\Psi $: 
\begin{eqnarray}
\del_a \ln \Psi = \frac{i}{\hbar } \del_a S \ \overset{\text{?}}{\sim}\  \frac{i}{\hbar }\del_a S^{\rm HJ} = \frac{i \pi_a }{\hbar }  \ \ \ \ \ \ \ \  (\hbar \rightarrow 0, \ x \rightarrow x_0 ) . 
\end{eqnarray}

Let us define outgoing and ingoing modes. In particle theory, we admit the future-directed condition (\ref{FDcond}), or $P(r) \ge 0$ outside the event horizon. However, for classical fields, there exist no restrictions on $P(r)$. We should define outgoing and ingoing modes so that they appropriately correspond to those of particle theory. For that purpose, we define a field-theoretical ``contravariant momentum'' $U^a $ as 
\begin{eqnarray}
U^a \!&=&\! g^{ab} (- i \hbar \del_b \ln \Psi - e A_b ) \\
	&=&\! g^{ab} (\del_b S - e A_b ) . 
\end{eqnarray}
When the real part of $U^a $ behaves asymptotically as ${\rm Re}(U^a ) \sim \frac{dx^a }{d\lam }$ around a point $x_0 $, where $x^a = x^a (\lam ) $ is the orbit of a particle motion, then we call $\Psi $ a pure mode at the point. When 
\begin{eqnarray}
P(r)\, {\rm Re}( U^r ) \sim P(r) \frac{dr}{d\lam } > 0 , 
\end{eqnarray}
then $\Psi $ is a pure outgoing mode at the point. Similarly, when 
\begin{eqnarray}
P(r)\, {\rm Re}( U^r ) \sim P(r) \frac{dr}{d\lam } < 0 , 
\end{eqnarray}
then $\Psi $ is a pure ingoing mode at the point. 

\ \\

For Kerr-Newman spacetime in Boyer-Lindquist coordinates, when 
\begin{eqnarray}
{\rm Re}\big( - i \hbar P(r) \del_r \ln \Psi \big) \sim P(r) \pi_r > 0 , \label{defoutgoingKN}
\end{eqnarray}
then $\Psi $ is a pure outgoing mode at $x_0 $, and when 
\begin{eqnarray}
{\rm Re}\big( - i \hbar P(r) \del_r \ln \Psi \big) \sim P(r) \pi_r < 0 , \label{defingoingKN}
\end{eqnarray}
then it is a pure ingoing mode at $x_0 $.

\subsection{The Klein-Gordon equation in Kerr-Newman spacetime}
We assume the scalar field $\Psi $ is written in the following form: 
\begin{eqnarray}
\Psi = f(r)g(\theta ) \exp\Big[ \frac{i}{\hbar } (-\eps t + m\phi ) \Big] . \label{decomp}
\end{eqnarray}
Substituting Eq.\,(\ref{decomp}) into Eq.\,(\ref{KG}), we can separate the Klein-Gordon equation in terms of $r$ and $\theta $~\cite{BMV2014}: 
\begin{eqnarray}
&\displaystyle\frac{d^2 f}{dr^2 } + \left( \frac{ 1 }{r - r_H } + \frac{ 1 }{r - r_C } \right)\frac{df}{dr} + \frac{1}{\hbar^2 \Delta^2 } \Big[ P(r)^2 - \Delta ( \mu^2 r^2 + \lam ) \Big] f = 0, &\label{sep1} \\
&\displaystyle\frac{\hbar^2 }{\si\theta } \frac{d}{d\theta } \bigg( \si\theta \frac{dg}{d\theta } \bigg) - \bigg[ \mu^2 a^2 \cos^2 \!\theta - \lam + \bigg( \eps a \si\theta - \frac{m}{\si\theta } \bigg)^2\, \bigg] g = 0, & \label{sep2}
\end{eqnarray}
where $\lam$ is the separation constant and $P(r)$ is defined by Eq.\,(\ref{defP}). 

The angular part of Eq.\,(\ref{sep2}) is a kind of spheroidal equation, which in a category of the confluent Heun equation. The solutions of the angular part of Eq.\,(\ref{sep2}) are the oblate spheroidal harmonic functions and $\lam$ are their eigenvalues, which are parameterized by two integers $m$ and $l$ such that $|m|\le l$, as is the case with the spherical harmonics~\cite{BCPFI1972}. However, the values of $\lam$ cannot be analytically expressed in terms of $m$ and $l$. In our case, $\lam$ takes on a real value.\footnote{In general, the eigenvalues of spin-weighted spheroidal harmonics are complex. They take on real values only in the oblate or prolate case with spin $0$~\cite{BCC2006}. } 

To see the counterpart of $\lam$ in particle theory, substituting $\displaystyle g(\theta ) = \exp\Big[ \frac{i}{\hbar} S_{\theta}(\theta)\Big]$ into Eq.\,(\ref{sep2}), we obtain 
\begin{eqnarray}
\lam = \bigg( \frac{dS_{\theta}}{d\theta} \bigg) ^2 + \mu^2 a^2 \cos^2 \!\theta + \frac{1}{\s2\theta} (m - a \eps \s2\theta )^2 - i \hbar \bigg[ \frac{\cos \theta}{\sin \theta} \frac{dS_{\theta}}{d\theta} + \frac{d^2 S_{\theta}}{d\theta^2 } \bigg] . \label{lam}
\end{eqnarray}
Comparing Eq.\,(\ref{Ktheta}) with Eq.\,(\ref{lam}), we find that $\lam$ is a field-theoretical counterpart of $\mathcal{K}$. 

As we will see below, one finds that the radial part of Eq.\,(\ref{sep1}) can also be written in the from of the confluent Heun equation and, in the extremal case Eq.\,(\ref{sep1}) is the double confluent Heun equation~\cite{BCV2014}~\cite{BMV2014}. Furthermore, in several special cases, Eq.\,(\ref{sep1}) reduces to essentially hypergeometric equations. 

\subsection{Local solutions in the nonextremal case}
The radial part of the Klein-Gordon equation (\ref{sep1}) is arranged in the form 
\begin{eqnarray}
&&\frac{d^2 f}{dr^2 } + \left( \frac{A_1 }{r - r_H } + \frac{A_2 }{r - r_C } + E_0 \right) \frac{df}{dr} \nonumber \\
	&&\ \ \ \ \ \ \ \ + \left[ \frac{B_1 }{(r - r_H )^2 } + \frac{B_2 }{(r - r_C )^2 } + \frac{C_1 }{r - r_H } + \frac{C_2 }{r - r_C } + D_0 \right] f = 0 ,  \label{ngCHE}
\end{eqnarray}
where 
\begin{eqnarray}
A_1 \!&=&\! A_2 \, =\, 1 , \ \ \ \ \ \  E_0 \, =\, 0 , \\
B_1 \!&=&\! \frac{P_H ^2 }{\hbar^2 (r_H - r_C )^2 } , \\
B_2 \!&=&\! \frac{P_C^2 }{\hbar^2 (r_H - r_C )^2 } , \\
C_1 \!&=&\! - \frac{2 P_H ^2 }{\hbar^2 (r_H - r_C )^3 } + \frac{2P^{\prime }_H P_H }{\hbar^2 (r_H - r_C)^2 } - \frac{\mu^2 r_H ^2 + \lam }{\hbar^2 (r_H - r_C )} , \label{defC1} \\
C_2 \!&=&\! \frac{2 P_C ^2 }{\hbar^2 (r_H - r_C )^3 } + \frac{2P^{\prime }_C P_C }{\hbar^2 (r_H - r_C)^2 } + \frac{\mu^2 r_C ^2 + \lam }{\hbar^2 (r_H - r_C )} , \label{defC2} \\
D_0 \!&=&\! \frac{\eps^2 - \mu^2 }{\hbar^2 } . 
\end{eqnarray}
With some exceptions, this equation has two regular singular points at $r = r_H, \ r_C$ and an irregular singular point at $r = \infty$. In general, second-order linear ordinary differential equations that have four regular singular points are called Heun equations, and those that have two regular singular points and an irregular singular point that is obtained by a confluent process of two regular singular points are called confluent Heun equations. The form of Eq.\,(\ref{ngCHE}) is called the {\it natural general form} of the confluent Heun equation~\cite{HDE}. An important exception is the case in which the irregular singular point at infinity becomes a regular singular point. We shall see that case in section \ref{Reduction}. 

Consider the following transformations of variables: 
\begin{eqnarray}
x \!&=&\! \frac{r - r_H }{r_C - r_H} , \\
f(r) \!&=&\! e^{\frac{1}{2} \alpha x } x^{\frac{1}{2} \beta } (x - 1)^{\frac{1}{2} \gamma } H(x) , \label{s-hom}
\end{eqnarray}
where 
\begin{eqnarray}
\alpha^2 \!&=&\! - 4 (r_H - r_C )^2 D_0 , \\
\beta^2 \!&=&\! - 4 B_1 , \label{beta2}  \\
\gamma^2 \!&=&\! - 4 B_2 . 
\end{eqnarray}

Equation (\ref{ngCHE}) transforms into 
\begin{eqnarray}
\frac{d^2 H}{dx^2} + \bigg( \alpha + \frac{\beta + 1}{x} + \frac{\gamma +1}{x - 1} \bigg) \frac{dH}{dx} + \bigg( \frac{\sig}{x} + \frac{\nu}{x - 1} \bigg) H = 0  , \label{cCHE}
\end{eqnarray}
where 
\begin{eqnarray}
\sig \!&=&\! \frac{1}{2} (\beta + 1)(\alpha - \gamma - 1) + \frac{1}{2} - \eta , \\
\nu \!&=&\! \frac{1}{2} (\gamma + 1)(\alpha + \beta + 1) - \frac{1}{2} + \delta + \eta , \\
\delta \!&=&\! - (r_H - r_C ) (C_1 + C_2 ) + \frac{1}{2} , \\
\eta \!&=&\! (r_H - r_C ) C_1 . 
\end{eqnarray}
The form of Eq.\,(\ref{cCHE}) is called the {\it nonsymmetrical canonical form} of the confluent Heun equation.\footnote{Our notation is slightly different from ~\cite{HDE} but is the same as ~\cite{BMV2014}, replacing $\mu $ by $\sig $. } A solution of Eq.\,(\ref{cCHE}) that is regular at $x = 0$ and whose value at $x = 0$ is $1$ can be obtained by a power series. Such a solution is called a local Frobenius solution of Eq.\,(\ref{cCHE}) around $x = 0$ or merely a local solution. Similarly, local Frobenius solutions around the other regular singular point $x = 1$ can also be obtained. Note that the local Frobenius solutions are ``local'' in the following two senses: First, a local Frobenius solution around a regular singular point, say $x = 0$, is not always that of the other regular singular point $x = 1$. Solutions that are Frobenius solutions around both the regular singular points are called confluent Heun functions. Given parameters of the equation, confluent Heun functions do not always exist.  

Second, since a local Frobenius solution is defined by a power series, the domain of definition is only inside its radius of convergence. In general, the domain of definition does not contain the other singular points. When you want global information on the local Frobenius solution, you need its global analytic continuation. However, that is unknown except for some special cases. Obvious exceptional cases are when the equations can reduce to essentially hypergeometric equations, each of which has only three regular singular points, or their confluent types. 

In this section, we see the local Frobenius solutions at $x = 0$, which are important for discussing the qualitative properties of the BSW effect. 

In nonextremal case $r_H \ne r_C $, substituting 
\begin{eqnarray}
H = Hl(\alpha, \beta, \gamma, \delta, \eta; x) = \sum_{n=0}^{\infty} c_n x^n 
\end{eqnarray}
into Eq.\,(\ref{cCHE}), one can obtain the 3-term recurrence relation\footnote{When $\beta $ is a negative integer, the recurrence relation  (\ref{recurC}) is not valid. In such a situation, the Frobenius solution contains a logarithmic term.}
\begin{eqnarray}
(n + 1) (n + 1 + \beta ) c_{n+1} + \left[ - n (n + 1 -\alpha + \beta + \gamma ) + \sig \right] c_n \nonumber \\
 + \left[ (n - 1)\alpha + \sig + \nu \right] c_{n-1} = 0  \label{recurC}
\end{eqnarray}
and the initial condition
\begin{eqnarray}
c_{-1} = 0 , \ \ \  c_0 = 1 . 
\end{eqnarray}
Here, $Hl(\alpha, \beta, \gamma, \delta, \eta; x)$ means the local Frobenius solution at $x = 0$. For large $n$, in general 
\begin{eqnarray}
\frac{c_n}{c_{n+1}} = 1 + \mathcal{O}\bigg( \frac{1}{n} \bigg) , 
\end{eqnarray}
so that the radius of convergence of the series is equal to unity. 

Substituting $n = 0$ into Eq.\,(\ref{recurC}), we obtain 
\begin{eqnarray}
c_1 = - \frac{\sig}{\beta + 1} . 
\end{eqnarray}

Thus, by Eq.\,(\ref{s-hom}) and the arbitrariness of the sign of $\beta $ in Eq.\,(\ref{beta2}), two local solutions of the radial part of the Klein-Gordon equation (\ref{ngCHE}) are obtained:  
\begin{eqnarray}
f^{out} (x) \!&=&\! e^{\frac{1}{2} \alpha x} (x - 1)^{\frac{1}{2} \gamma } x^{\frac{1}{2} \beta } Hl(\alpha , \beta , \gamma , \delta , \eta ; x) , \label{fononext} \\ 
f^{in} (x) \!&=&\! e^{\frac{1}{2} \alpha x} (x - 1)^{\frac{1}{2} \gamma } x^{-\frac{1}{2} \beta } Hl(\alpha , -\beta , \gamma , \delta , \eta ; x) , \label{finonext}
\end{eqnarray}
where explicitly we choose 
\begin{eqnarray}
\alpha \!&=&\! 2 \hbar^{-1} (r_H - r_C) \sqrt{\mu^2 - \eps^2 } , \\
\beta \!&=&\! \frac{2 i P_H }{\hbar (r_H - r_C )} , \label{defbeta} \\
\gamma \!&=&\! \frac{2 i P_C }{\hbar (r_H - r_C )} , \\
\delta \!&=&\! \hbar^{-2} [ - 2 \eps (P_H - P_C ) + \mu^2 (r_H - r_C )(r_H + r_C ) ] , \\
\eta \!&=&\! \hbar^{-2} \bigg[ - \frac{2P_H ^2 }{(r_H - r_C )^2 } + \frac{2P_H ^{\prime } P_H }{r_H - r_C } - (r_H^2 \mu ^2 + \lam ) \bigg] . 
\end{eqnarray}
One can check that $f^{out} (x)$ and $f^{in} (x)$ are pure outgoing and ingoing solutions respectively at the event horizon, by substituting into Eq.\,(\ref{defoutgoingKN}) and Eq.\,(\ref{defingoingKN}): 
\begin{eqnarray}
- i \hbar P(r) \del_r \ln \Psi^{out} \!&=&\! - i \hbar P(r) \bigg[ \frac{\beta }{2(r - r_H)} + \frac{\gamma }{2(r - r_C)} \nonumber \\
	&& - \frac{\alpha }{2(r_H - r_C) } - \frac{\del_x \ln Hl(\alpha , \beta , \gamma , \delta , \eta ; x )}{r_H - r_C} \bigg] \\
	&\sim & \frac{P_H ^2 }{(r_H - r_C ) (r - r_H )} \\
	&\sim & \left| P_H \frac{\sqrt{R}}{\Delta } \right| \\
	&=& \left| P_H \pi_r \right| \ \ge \ 0 , \\
- i \hbar P(r) \del_r \ln \Psi^{in} \!&\sim &\! - \frac{P_H ^2 }{(r_H - r_C ) (r - r_H )} \\
	&\sim & - \left| P_H \pi_r \right| \ \le \  0 . 
\end{eqnarray}

\subsection{Asymptotic expansions in the extremal case}
In the extremal case $r_H = r_C $,  the two regular singular points of Eq.\,(\ref{ngCHE}) coincide and the equation has two irregular singular points. It is implied that the radial part of the Klein-Gordon equation can be written as the double confluent Heun equation. 

Indeed, under the transformation of the variable 
\begin{eqnarray}
\xi = r - M , 
\end{eqnarray}
one can arrange the radial equation (\ref{sep1}) in the form 
\begin{eqnarray}
\xi^2 \frac{d^2 f}{d\xi ^2} + 2 \xi \frac{df}{d\xi} + \sum_{i=-2}^{2} b_i \xi ^i f = 0 ,  \label{DCHE}
\end{eqnarray}
where 
\begin{eqnarray}
b_2 \!&=&\! \hbar^{-2} \left( \eps^2 - \mu^2 \right) , \\
b_1 \!&=&\! \hbar^{-2} \left( P^{\prime \prime } _H P^{\prime } _H - 2 M \mu ^2 \right) \\
	\!&=&\! \hbar^{-2} \left[ 2M (2\eps^2 - \mu^2 ) - 2 \eps eQ \right] , \\
b_0 \!&=&\! \hbar^{-2} \left[ P^{\prime \prime }_H P_H + (P^{\prime }_H )^2 - (M^2 \mu^2 + \lam ) \right] \\
	&=&\! \hbar^{-2} \left[ 6M \eps (M \eps - Q e ) + 2a \eps (a \eps - m ) + Q^2 e^2 - M^2 \mu^2 - \lam \right] , \\
b_{-1} \!&=&\! 2\hbar^{-2} P^{\prime }_H P_H \\
	\!&=&\! 2\hbar^{-2} (2M\eps - eQ ) \left[ (M^2 + a^2 ) \eps - am - eQM \right] , \\
b_{-2} \!&=&\! \hbar^{-2} P_H ^2 \\
	\!&=&\!\hbar^{-2} \left[ (M^2 + a^2 )\eps - am -eQM \right]^2 . 
\end{eqnarray}
The form of Eq.\,(\ref{DCHE}) is the {\it general double confluent Heun equation} with $a_1 = a_{-1} = 0$ and $a_0 = 1$ in~\cite{HDE}. 

In a procedure that is similar to that for the confluent Heun equation, one can formally obtain power series around an irregular singular point, say $\xi = 0$. However, the radii of convergence of the series are $0$. Therefore, such series do not specify analytic functions, but merely imply the existence of solutions that admit the series as asymptotic expansions. 

Let us define the following quantities\footnote{The signs are chosen so as to be consistent to the physical terms ``outgoing'' and ``ingoing''. } 
\begin{eqnarray}
\alpha_1 \!&=&\! 2 \sqrt{- b_2} \ =\  2 \hbar^{-1} \sqrt{\mu^2 - \eps^2 } , \\
\alpha_{-1} \!&=&\! \mathrm{sign}(P_H ) \, 2 i \hbar^{-1} \sqrt{b_{-2}} \ =\  2 i \hbar^{-1} P_H , \\
\beta_{-1} \!&=&\! \frac{b_{-1}}{\alpha_{-1}} \ =\  - i \hbar^{-1} P^{\prime }_H . 
\end{eqnarray}
It is known that there exist two solutions $f^{out} (\xi )$ and $f^{in} (\xi )$ that admit the following asymptotic representations~\cite{HDE}: 
\begin{eqnarray}
f^{in} (\xi) \!&\sim &\! \xi^{\beta_{-1}} \exp \bigg( \frac{\alpha_1 }{2} \xi + \frac{\alpha_{-1} }{2\xi} \bigg) \sum_{n=0}^{\infty} \psi_n ^+ \bigg( \frac{\xi}{\alpha_{-1}} \bigg)^{n} , \label{foext} \\
f^{out} (\xi) \!&\sim &\! \xi^{-\beta_{-1}} \exp \bigg( \frac{\alpha_1 }{2} \xi - \frac{\alpha_{-1} }{2\xi} \bigg) \sum_{n=0}^{\infty} \psi_n ^- \bigg( \frac{\xi}{- \alpha_{-1}} \bigg)^{n} . \label{fiext} 
\end{eqnarray}
Here, the coefficients $\psi_n ^{\pm }$ are given by the 3-term recurrence relation
\begin{eqnarray}
(n+1)\psi_{n+1} ^{\pm } - \bigg[\Big(n \pm \beta_{-1} + \frac{1}{2}\Big)^2 + b_0 - \frac{1}{4} \mp \frac{\alpha_1 \alpha_{-1}}{2} \bigg]\psi_n ^{\pm } \nonumber \\
\mp \alpha_1 \alpha_{-1} \bigg( n \pm \beta_{-1} + \frac{b_1}{\alpha_1} \bigg) \psi_{n-1} ^{\pm } = 0 , 
\end{eqnarray}
where double-signs apply in the same order, and the initial condition
\begin{eqnarray}
\psi_{-1} ^{\pm } = 0 ,  \ \ \ \psi_0 ^{\pm } = 1 . 
\end{eqnarray}
Substituting $n = 0$, we obtain 
\begin{eqnarray}
\psi_1^{\pm } = \bigg( \pm \beta_{-1} + \frac{1}{2} \bigg)^2 + b_0 - \frac{1}{4} \mp \frac{\alpha_1 \alpha_{-1} }{2} . \label{psipm}
\end{eqnarray}

\section{Reductions to (confluent) hypergeometric equations}\label{Reduction}
In this section, we study the special cases in which the confluent Heun equation\,(\ref{ngCHE}) and the double confluent Heun equation\,(\ref{DCHE}) reduce to the hypergeometric equation or its confluent type. 

In this section, for simplicity, we adopt the unit $\hbar = 1$. 

\subsection{Nonextremal case for specific marginal modes}\label{Nonextmarginal}
In generic parameters, Eq.\,(\ref{ngCHE}) has an irregular singular point at infinity. To see clearly around infinity, by the transformation of the variable 
\begin{eqnarray}
w = \frac{r_C - r_H }{r - r_H }, 
\end{eqnarray}
Eq.\,(\ref{ngCHE}) is transformed to 
\begin{eqnarray}
&&\frac{d^2 f}{dw^2 } + \left( \frac{2 - A_1 - A_2 }{w} + \frac{A_2 }{w - 1} - \frac{E_0 }{w^2 } \right) \frac{df}{dw} \nonumber \\
	&&\ \ \ \ \ \ \ \  + \left( \frac{B_1 }{w^2 } + \frac{B_2 }{w^2 (w - 1 )^2 } + \frac{C_1 }{w^3 } - \frac{C_2 }{w^3 (w - 1)} + \frac{D_0 }{w^4 } \right) f = 0 .   
\end{eqnarray}
The conditions for the infinity $w = 0$ to be a regular singular point are 
\begin{eqnarray}
C_1 + C_2 = 0  \ \ \ \ \ \ {\rm and} \ \ \ \ \ D_0 = E_0 = 0 .  \label{marginal1CDE}
\end{eqnarray}
From Eqs.\,(\ref{defC1}) and (\ref{defC2}), and some algebra calculations, 
\begin{eqnarray}
C_1 + C_2 = 2\eps (2\eps M - e Q ) - 2 M \mu^2 . 
\end{eqnarray}
Combining with $D_0 = \eps^2 - \mu^2 = 0$, condition (\ref{marginal1CDE}) is equivalent to\footnote{Note that condition (\ref{marginal1}) is equivalent to that of ``case-1'' in Section III of~\cite{WC1999}.}  
\begin{eqnarray}
\eps^2 = \mu^2 \ \ \ \ {\rm and}\ \ \ \ \eps (\eps M - eQ) = 0 . \label{marginal1}
\end{eqnarray}
The condition $\eps^2 = \mu^2 $ means that the configurations represent marginally bound states. That is why we call the modes that satisfy $\eps^2 = \mu^2 $ the ``marginal modes''. 

In fact, in condition (\ref{marginal1CDE}), Eq.\,(\ref{ngCHE}) is transformed to the canonical form of the hypergeometric equation 
\begin{eqnarray}
x(1 - x ) \frac{d^2 \hat{f}}{dx^2 } + \left[ c - (a + b + 1) x \right] \frac{d\hat{f}}{dx} - ab \hat{f} = 0 , 
\end{eqnarray}
where 
\begin{eqnarray}
a + b \!&=&\! 1 + 2 \Lambda_1 + 2 \Lambda_2 , \label{defa+b} \\
a b \!&=&\! 2 \Lambda_1 \Lambda_2 + \Lambda_1 + \Lambda_2 + (r_H - r_C )C_1 , \label{defab} \\
c \!&=&\! 1 + 2 \Lambda_1 ,  \label{defcLam} \\
( \Lambda_1 ) ^2 \!&=&\! - B_1 , \label{defLam1} \\
(\Lambda_2 ) ^2 \!&=&\! - B_2 ,  \label{defLam2} 
\end{eqnarray}
by the transformations of the variables 
\begin{eqnarray}
x = \frac{r - r_H }{r_C - r_H } , \ \ \ \ \ \ \ \ f(r) = x^{\Lambda_1 } (1 - x)^{\Lambda_2 } \hat{f}(x) . 
\end{eqnarray}

The definitions of $\Lambda_1 $ and $\Lambda_2 $, namely Eqs.\,(\ref{defLam1}) and (\ref{defLam2}), have arbitrariness of their signs. We choose them as 
\begin{eqnarray}
\Lambda_1 \!&=&\! - \frac{i P_H }{r_H - r_C } , \\
\Lambda_2 \!&=&\! \frac{i P_C }{r_H - r_C } . 
\end{eqnarray}
Therefore, two linearly independent solutions are obtained 
\begin{eqnarray}
f_1 \!&=&\! x^{\Lambda_1 } (1 - x)^{\Lambda_2 } F(a , b ; c ; x ) , \label{deff1} \\
f_2 \!&=&\! x^{- \Lambda_1 } (1 - x)^{\Lambda_2 } F(1 + a - c , 1 + b - c ; 2 - c ; x) , \label{deff2}
\end{eqnarray}
where $F(a, b ; c ; x)$ is the hypergeometric function defined by 
\begin{eqnarray}
F(a , b ; c ; x) = \sum_{n = 0}^{\infty} \frac{(a)_n (b)_n }{n!\, (c)_n } x^n  
\end{eqnarray}
and its analytic continuation. The symbol $(a)_n$ is the Pochhammer symbol defined by
\begin{eqnarray}
(a)_0 = 1 , \ \ \ \ (a)_n = a(a + 1)\dots (a + n -1) . 
\end{eqnarray}
From Eqs.\,(\ref{defa+b}), (\ref{defab}) and (\ref{defcLam}), 
\begin{eqnarray}
a \!&=&\! - i \frac{P_H - P_C }{r_H - r_C } + \frac{1}{2} \left\{ 1 - i \sqrt{4 \big[ B_1 + B_2 + (r_H - r_C ) C_1 \big] - 1} \right\} \\
	&=&\! - i (2M\eps - Q e) + \frac{1}{2} \left\{ 1 - i \sqrt{ 4\big[ 2\eps P_H + (2M\eps - Q e)^2 - r_H ^2 \mu^2 - \lam \big] - 1 } \right\} , \\
b \!&=&\! - i (2M\eps - Q e) + \frac{1}{2} \left\{ 1 + i \sqrt{ 4\big[ 2\eps P_H + (2M\eps - Q e)^2 - r_H ^2 \mu^2 - \lam \big] - 1 } \right\} , \\
c \!&=&\ 1 - \frac{2i P_H }{r_H - r_C } . \label{defc}
\end{eqnarray}

Using the transformation formula~\cite{HMF1964}, 
\begin{eqnarray}
F(a, b; c; z) \!&=&\! \frac{\Gamma(c) \Gamma(b - a) }{\Gamma(b) \Gamma(c - a)} (-z)^{-a} F(a, 1 - c + a ; 1 - b + a ; z^{-1} ) \nonumber \\
	&& + \frac{\Gamma(c) \Gamma(a - b) }{\Gamma(a) \Gamma(c - b)} (-z)^{-b} F(b, 1 - c + b ; 1 - a + b ; z^{-1} ) , \label{transformulaHG}
\end{eqnarray}
where $\Gamma (z)$ is the gamma function, which satisfies the relations $\Gamma (z + 1) = z \Gamma(z)$ and $\Gamma(1) = 1$, $f_1 $ and $f_2 $ are expressed as 
\begin{eqnarray}
f_1 \!\!\!&=&\!\!\! x^{\Lambda_1 } (1 - x)^{\Lambda_2 } \bigg[ \frac{\G(c) \G(b - a)}{\G(b) \G(c - a)} (- x)^{- a} F\big( a, 1 - c + a ; 1 + a - b ; x^{- 1} \big) \nonumber \\
		&& + \frac{\G(c) \G(a - b)}{\G(a) \G(c - b)} (- x)^{- b} F\big( b , 1 - c + b ; 1 - a + b ; x^{- 1} \big) \bigg] , \label{nonextf1inf} \\
f_2 \!\!\!&=&\!\!\! x^{- \Lambda_1 } (1 - x)^{\Lambda_2 } \bigg[ \frac{\G(2 - c) \G(b - a)}{\G(1 + b - c) \G(1 - a)} (- x)^{- 1 + c - a} F\big( 1 - c + a, a ; 1 + a - b ; x^{- 1} \big) \nonumber \\
		&& + \frac{\G(2 - c) \G(a - b)}{\G(1 - c + a) \G(1 - b)} (- x)^{- 1 - b + c} F\big( 1 + b - c , b ; 1 - a + b ; x^{- 1} \big) \bigg] . \label{nonextf2inf}
\end{eqnarray}

From Eq.\,(\ref{deff1}), the logarithmic derivative of $f_1 $ is 
\begin{eqnarray}
\del_r \ln f_1 \!&=&\! -\frac{1}{r_H - r_C } \left[ \frac{\Lambda_1 }{x} - \frac{\Lambda_2 }{1 - x} + \frac{a b}{c} \frac{F(a + 1 , b + 1 ; c + 1 ; x)}{F(a , b ; c ; x)} \right] \\
	&\sim&\! \frac{\Lambda_1 }{r - r_H } \, = \, - \frac{i P_H }{(r_H - r_C ) (r - r_H )}  \ \ \ \ \ \ \ \ \ {\rm (near\ the\ event\ horizon)}. 
\end{eqnarray}
From Eq.\,(\ref{pr}), the canonical momentum $\pi_r $ is 
\begin{eqnarray}
\pi_r \!&=&\! \sig_r \frac{\sqrt{R(r)}}{\Delta } \\
	&\sim &\! \sig_r \frac{P_H }{(r_H - r_C ) (r - r_H )} . \ \ \ \ \ \ \ \ \ {\rm (near\ the\ event\ horizon)}
\end{eqnarray}
One can find  
\begin{eqnarray}
- i P(r) \del_r \ln f_1 \!&\sim&\! P_H \pi_r (\sig_r = -1) \ \ \ \ \ \ \ \ {\rm (near\ the\ event\ horizon)} \\
	&\sim&\! - \frac{P_H ^2 }{(r_H - r_C ) (r - r_H )} \, \le 0 . 
\end{eqnarray}
Therefore, $f_1 $ is a pure ingoing solution at the event horizon. 

Similarly, 
\begin{eqnarray}
- i P(r) \del_r \ln f_2 \, \sim \, P_H \pi_r (\sig_r = 1) \, \ge \, 0   \ \ \ \ \ \ \ \ {\rm (near\ the\ event\ horizon)}, 
\end{eqnarray}
i.e., $f_2 $ is a pure outgoing solution at the event horizon. 

In the far region $r \gg r_H $, the logarithmic derivative of the first line of Eq.\,(\ref{nonextf1inf}) behaves as 
\begin{eqnarray}
\del_r \ln \Big( {\rm the\ first\ line\ of\ Eq.\,(\ref{nonextf1inf})} \Big) \, \sim \, \frac{1}{r} \left( - \frac{1}{2} + \frac{i}{2} \sqrt{4[B_1 + B_2 + (r_H - r_C ) C_1 ] - 1 } \right) . 
\end{eqnarray}
Now $\pi_r ^2 $ can be arranged in the form 
\begin{eqnarray}
\pi_r ^2 \!&=&\! \frac{R(r)}{\Delta ^2 } \\
	&=&\! D_0 + \frac{C_1 }{r - r_H } + \frac{C_2 }{r - r_C} + \frac{B_1 }{(r - r_H )^2 } + \frac{B_2 }{(r - r_C )^2 } \\
	&\simeq &\! D_0 + \frac{C_1 + C_2 }{r - r_H } + \frac{- (r_H - r_C ) C_2 + B_1 + B_2 }{(r - r_H )^2 } + \mathcal{O}\big( (r - r_H )^{-3} \big) . 
\end{eqnarray}
When $C_1 + C_2 = D_0 = 0$, 
\begin{eqnarray}
\pi_r ^2 \simeq \frac{B_1 + B_2 + (r_H - r_C ) C_1 }{(r - r_H )^2 } + \mathcal{O}\big( (r - r_H )^{- 3} \big) . 
\end{eqnarray}
Therefore, in the far region, 
\begin{eqnarray}
\del_r \ln \Big( {\rm the\ first\ line\ of\ Eq.\,(\ref{nonextf1inf})} \Big) \, \sim \, \frac{1}{r} \left( -\frac{1}{2} + \frac{i}{2} \sqrt{4 (r - r_H )^2 \pi_r ^2 - 1 } \right) . 
\end{eqnarray}
Writing $\hbar $ explicitly, 
\begin{eqnarray}
&& - i \hbar P(r) \del_r \ln \Big( {\rm the\ first\ line\ of\ Eq.\,(\ref{nonextf1inf})} \Big) \nonumber \\
	&\sim& - i \frac{P(r) }{r} \left( -\frac{\hbar }{2} + \frac{i}{2} \sqrt{4 (r - r_H )^2 \pi_r ^2 - \hbar ^2 } \right) \\
	&\sim& P(r) |\pi_r |   \ \ \ \ \ \ \ \ \ \ \ \ (\hbar \rightarrow 0 ,\ r \gg r_H ). 
\end{eqnarray}
Thus, when $P(r) > 0$, the first line of Eq.\,(\ref{nonextf1inf}) represents the outgoing part of $f_1 $ in the far region $r \gg r_H $. Similar calculations reveal that the second line of Eq.\,(\ref{nonextf1inf}) is the ingoing part in the far region.

We adopt the normalization factor $N$, which adjusts the amplitude of the ingoing part of $N f_1 (x_0 )$ at a distant point $x_0 $ to $C_0 ^{in} $. Explicitly, 
\begin{eqnarray}
N f_1 \!&\simeq &\! C_0 ^{in} (-1)^{\Lambda_1 } \frac{\G(a) \G(c - b)}{\G(c) \G(a - b)} (- x_0 )^{- A}  f_1 \label{Nf1} \\
	&\simeq &\! C_0 ^{in} \frac{\G(a) \G(c - b) \G(b - a)}{\G(b) \G(c - a) \G(a - b)} (- x_0 )^{B} \left( \frac{x}{x_0 } \right)^{C} F\big( a, 1 - c + a ; 1 - b + a ; x^{- 1} \big) \nonumber \\
		&& + C_0 ^{in} \bigg( \frac{x}{x_0 } \bigg)^A F\big( b , 1 - c + b ; 1 - a + b ; x^{- 1} \big) , \label{Nf1nonext} 
\end{eqnarray}
where 
\begin{eqnarray}
A \!&=&\! - \frac{1}{2} - \frac{i}{2} \sqrt{4\big[ 2\eps P_H + (2M\eps - Q e)^2 - r_H ^2 \mu^2 - \lam \big] - 1} , \\
B \!&=&\! i \sqrt{4\big[ 2\eps P_H + (2M\eps - Q e)^2 - r_H ^2 \mu^2 - \lam \big] - 1 } , \\
C \!&=&\! - \frac{1}{2} + \frac{i}{2} \sqrt{4\big[ 2\eps P_H + (2M\eps - Q e)^2 - r_H ^2 \mu^2 - \lam \big] - 1} . 
\end{eqnarray}

When $a = b + k $ and $k = 0,\, 1,\, 2, \dots $, then both the first and second terms of the right-hand side of Eq.\,(\ref{nonextf1inf}) diverge. In this case, the following formula~\cite{HMF1964} is useful: 
\begin{eqnarray}
F(b + k , b ; c ; z ) \!&=&\! \frac{\G(c) (- z)^{- b - k} }{\G(b + k) \G(c - b)} \sum_{n=0}^{\infty } \frac{(b)_{n + k} (1 - c + b)_{n + k} }{n! (n + k)! } z^{- n} [\ln(- z) \nonumber \\
	&&\!\!\! +\, \psi (1 + k + n) + \psi (1 + n) - \psi (b + k + n) - \psi (c - b - k - n)] \nonumber \\
	&&\!\!\! +\, (- z)^{- b} \frac{\G(c)}{\G(b + k)} \sum_{n = 0}^{k - 1} \frac{\G(k - n) (b)_n }{n! \G(c - b - n)} z^{- n} ,  
\end{eqnarray}
where $\psi (z)$ is the digamma function, which is defined as the logarithmic derivative of the gamma function: $\psi (z) = \Gamma^{\prime } (z) / \Gamma (z)$. For example, when $a = b$ (or $k = 0$), 
\begin{eqnarray}
f_1 \!&=&\! (-1)^{\Lambda_1 } \frac{\G(c)}{\G(a) \G(c - a)} (- x)^{- \frac{1}{2}} \sum_{n = 0} ^{\infty } \frac{(a)_n (1 + a - c)_n }{(n!)^2 } x^{- n} \nonumber \\
	&& \times [\ln (- x) + 2 \psi (n + 1) - \psi (a + n) - \psi (c - a - n)] . \label{f1a=b}
\end{eqnarray}
This solution behaves $\sim x^{- \frac{1}{2}} $ in the far region. 

Note that when $a = b$, the first line of Eq.\,(\ref{Nf1nonext}) coincides with the second line. In other words, when $a = b$, the outgoing part and the ingoing part degenerate in the far region, so that our normalization procedure becomes invalid.

\subsubsection{The general Legendre equation}\label{gLegendre}
With additional conditions, the radial equation becomes somewhat simpler. For that purpose, it is easier to see Eq.\,(\ref{ngCHE}) in another form. In the transformation of variable 
\begin{eqnarray}
u = \frac{2r - 2M}{r_H - r_C } , 
\end{eqnarray}
then Eq.\,(\ref{ngCHE}) can be arranged in the form of the {\it generalized spheroidal equation}~\cite{HDE}: 
\begin{eqnarray}
\frac{d}{du} \left[ (u^2 - 1 ) \frac{d}{du} \right] f + \left( - p^2 (u^2 - 1) + 2p\beta u - \nu (\nu +1) - \frac{n^2 + s^2 + 2ns u }{u^2 - 1}  \right) f = 0 , \label{GSE}
\end{eqnarray}
where using $A_1 = A_2 = 1$ and $E_0 = 0$, 
\begin{eqnarray}
&p^2 & = \  - \frac{(r_H - r_C )^2 }{4} D_0 , \\
&(n + s)^2 &=\  -4B_1 , \\ 
&(n - s)^2 &= \ -4B_2 , \\ 
&2p\beta & = \ \frac{r_H - r_C }{2} (C_1 + C_2 ) , \\ 
&\nu (\nu + 1)& =\  \frac{r_H - r_C }{2} (C_2 - C_1 ) - (B_1 + B_2 ) .  
\end{eqnarray}
If the conditions (\ref{marginal1CDE}), or (\ref{marginal1}), are satisfied, then $p = \beta = 0$ so that Eq.\,(\ref{GSE}) is the same form as the angular part of the spin-weighted {\it spherical} wave equation. 

If, besides condition (\ref{marginal1CDE}), $B_1 = B_2$ is satisfied, then $s = 0$ (or $n = 0$) so that Eq.\,(\ref{GSE}) reduces to the {\it general Legendre equation}. Conditions (\ref{marginal1CDE}) and $B_1 = B_2 $ are satisfied when\footnote{It was pointed out in~\cite{WC1999} that in the situation (\ref{condL}), the solutions degenerate to be Legendre functions.} 
\begin{eqnarray}
&{\rm (i) }&\eps^2 = \mu^2 = e Q =0 , \ \ \ \ \ \ \ \ \ \ \ \ \ \, \Rightarrow \ P_H = P_C = - am , \label{marginal2a} \label{condL}\\
&{\rm (ii) }& \eps^2 = \mu^2 = 0 , \ \ \ e Q = - \frac{a m}{M} , \ \ \Rightarrow \ P_H = - P_C = \frac{am}{M}(r_H - M) , \label{marginal2b} \\
&{\rm (iii) }& \eps^2 = \mu^2 \ne 0, \ \ \eps M = e Q ,  \ \ a m = (M^2 - Q^2 ) \eps , \nonumber \\
	&&\ \ \ \ \ \ \ \ \ \ \ \ \ \ \ \ \ \ \ \ \ \ \ \ \ \ \ \ \ \ \ \ \ \ \ \ \ \ \Rightarrow \ P_H = - P_C = \eps M (r_H - M) . \label{marginal2c}
\end{eqnarray}
In these cases, Eq.\,(\ref{GSE}) is explicitly 
\begin{eqnarray}
\frac{d}{du} \left[ (u^2 - 1 ) \frac{d}{du} \right] f + \left( -\nu (\nu + 1) - \frac{ n^2 }{u^2 - 1} \right) f = 0 , \label{GLE}
\end{eqnarray}
where  
\begin{eqnarray}
n = \frac{2 i P_H }{r_H - r_C } 
\end{eqnarray}
and 
\begin{eqnarray}
\nu \!&=&\! \left\{ 
	\begin{array}{l}
	- \frac{1}{2} \Big[ 1 + i \sqrt{- 4\lam - 1}  \Big] \ \ \ \ \ \ \ \ \ \ \ \ \ \ \ \ \ \ \ \ \ \ \ \ \ \ \ \ \ \ \ \ \ \ \  {\rm case\ (i) },  \vspace{2mm}\\
	- \frac{1}{2} \Big[ 1 + i \sqrt{4 Q^2 e^2 - 4\lam - 1}  \Big] \ \ \ \ \ \ \ \ \ \ \ \ \ \ \ \ \ \ \ \ \ \ \ \ \ \, {\rm case\ (ii) },  \vspace{2mm}\\
	- \frac{1}{2} \Big[ 1 + i \sqrt{- 4 (M^2 - a^2 - Q^2 )\eps^2 - 4\lam - 1}  \Big] . \ \ \ \ \  {\rm case\ (iii) }.  
	\end{array}
	\right.
\end{eqnarray}
Since neither $n$ nor $\nu $ is restricted to being integer, Eq.\,(\ref{GLE}) is the {\it general Legendre equation}.

The solutions of the  general Legendre equation are known as the general Legendre functions of the first kind $P_{\nu }^n (u)$ and those of the second kind $Q_{\nu }^n (u)$, which are essentially hypergeometric functions~\cite{HMF1964}: 
\begin{eqnarray}
f_P^+ (u) \!&=&\! P_{\nu }^n (u) \nonumber \\
	\!&=&\! \frac{1}{\Gamma(1 - n)} \left( \frac{1 + u}{1 - u} \right)^{ \frac{n}{2}} F\big( - \nu , \nu + 1; 1 - n ; {\textstyle\frac{1 - u}{2} } \big) , \\ 
f_Q (u) \!&=&\! Q_{\nu }^n (u) \nonumber \\
	\!&=&\! \frac{\sqrt{\pi }\, \Gamma \big( \nu + n + 1\big) }{2^{\nu + 1} \Gamma \big( \nu + \frac{3}{2} \big) } \frac{e^{i n \pi } (u^2 - 1)^{- \frac{n}{2}} }{u^{\nu + n +1 } } \nonumber \\
	&& \times \, F\Big( \frac{ \nu + n + 1}{2} , \frac{ \nu + n + 2}{2} ; \nu + \textstyle\frac{3}{2} ; u^{-2} \Big) .  
\end{eqnarray}
The function $f_P^- (u)$ obtained through replacing $n$ by $-n$ in $f_P^+ (u)$ is also a solution of Eq.\,(\ref{GLE}).\footnote{Similarly, the function obtained by replacing $n$ with $-n$ in $f_Q (u)$ is also a solution of Eq.\,(\ref{GLE}), but it is a constant multiple of $f_Q(u) $ itself. } Each $f_P^+ (u)$ represents a solution that contains a pure ingoing mode at the event horizon $u = 1$, and similarly $f_P^- (u)$ for outgoing. Using a transformation formula for hypergeometric functions, one can find that the solution $f_Q (u)$ describes a superposition of ingoing and outgoing mode at the event horizon.  

Using the transformation formula (\ref{transformulaHG}), the asymptotic behavior of $f_P^+ (u)$ at infinity is written as  
\begin{eqnarray}
f_P^+ (u) \!\!\!\!\!&=&\!\!\!\!\! \left( \frac{1 + u}{1 - u} \right)^{\frac{n}{2}} \!\! \bigg[ \frac{\Gamma(2\nu + 1)}{\Gamma(\nu + 1) \Gamma(\nu + 1 - n)} \left( \frac{u - 1}{2} \right)^{\nu } F\big( \textstyle-\nu , n - \nu ; -2\nu ; \frac{2}{1 - u} \big)   \nonumber \\
	&& + \frac{\Gamma(-2\nu - 1)}{\Gamma(-\nu) \Gamma(-\nu - n)} \left( \frac{u - 1}{2} \right)^{-\nu - 1} F\big( \nu + 1 , \nu + 1 + n ; 2\nu + 2 ; \textstyle \frac{2}{1 - u} \big) \bigg] \label{fPinfty} \\
\!&\sim&\! \frac{\Gamma(1 + 2 \nu )}{\Gamma(\nu + 1) \Gamma(1 - n + \nu ) } \left( \frac{u}{2} \right)^{\nu }  +  \frac{\Gamma(-1 - 2 \nu )}{\Gamma( -\nu ) \Gamma( - n - \nu ) } \left( \frac{u}{2} \right)^{-\nu - 1 } . \label{asymfP}
\end{eqnarray}
The solution $f_P^+ (u)$ represents a superposition of ingoing and outgoing modes around infinity. In fact, e.g., in case (i), 
\begin{eqnarray}
\left( \frac{u}{2} \right)^{\nu} \!&=&\! \left( \frac{u}{2} \right)^{-\frac{1}{2}} \left( \frac{u}{2} \right)^{-\frac{ i \sqrt{-1 - 4\lam }}{2}} \\
	\!&=&\! \left( \frac{r - M}{r_H - r_C } \right)^{-\frac{1}{2}} \exp \left( - i \int \frac{\sqrt{-1 - 4\lam }}{2(r - M)} dr \right) \ \ \ \ \ (-1 - 4\lam \ge 0) \\
	&=&\!  \left( \frac{r - M}{r_H - r_C } \right)^{-\frac{1}{2}} \exp \left( \int \frac{\sqrt{1 + 4\lam }}{2(r - M)} dr \right) \ \ \ \ \ \ \ \ \ \ (-1 - 4\lam < 0) ,
\end{eqnarray}
so when $-1 - 4\lam > 0$, the first term of Eq.\,(\ref{asymfP}) is an ingoing mode of wavelength $\sim \frac{2r}{\sqrt{-1 - 4\lam }}$. When $-1 - 4\lam < 0$, it blows up. 

One can identify a normalized solution $N f^{in}(u)$ such that $N f^{in}(u)$ contains a pure ingoing mode at the event horizon $u = 1$ and the ingoing amplitude at a distant point $r = r_0 \gg M$ is $C_0^{in} $: 
\begin{eqnarray}
N f^{in} (u) \!&=&\! C_0^{in} \frac{\Gamma(\nu + 1) \Gamma(\nu - n + 1)}{\Gamma(2\nu + 1)} \left( \frac{r_0 }{r_H - r_C } \right)^{- \nu } P_{\nu}^{n} (u) \\
	\!&=&\! C_0^{in} \frac{\Gamma(\nu + 1) \Gamma(\nu - n + 1)}{\Gamma(1 - n) \Gamma(2\nu + 1)} \left( \frac{r_0 }{r_H - r_C } \right)^{ - \nu } \nonumber \\ 
	&&\ \ \ \ \ \  \times  \left( \frac{1 + u}{1 - u} \right)^{\frac{n}{2}} F \big( -\nu , \nu + 1 ; 1 - n ; \textstyle\frac{1 - u}{2} \big) . 
\end{eqnarray}

\subsection{Extremal case}
The double confluent Heun equation (\ref{DCHE}) reduces to the confluent hypergeometric equation in the following two cases: 
\begin{eqnarray}
{\rm (i) }\  b_{-2} = b_{-1} = 0  \label{critcond} 
\end{eqnarray}
or 
\begin{eqnarray}
{\rm (ii) }\  b_2 = b_1 = 0 . \label{margcond}
\end{eqnarray}
The first condition (\ref{critcond}) is satisfied in the case\footnote{This case is contained in~\cite{Hod2014b} as the extremal limit of the far region. } of critical mode $P_H = 0$. The second condition (\ref{margcond}) represents marginal modes with specific charges, i.e., 
\begin{eqnarray}
\eps^2 = \mu^2  \ \ \ \text{and} \ \ \ \eps (M \eps - Q e ) = 0 . 
\end{eqnarray}
Therefore, in these two cases we obtain the global representations of the solutions by means of confluent hypergeometric functions. 

In section \ref{secCHGE}, we derive some formulae for the confluent hypergeometric equation. In section \ref{secextcrit}, we apply them to case (i), and in section \ref{secmarginal}, to case (ii).

\subsubsection{The confluent hypergeometric equation}\label{secCHGE}
Before dealing with specific cases (\ref{critcond}) and (\ref{margcond}), we shall derive some formulae for the general confluent hypergeometric equation: 
\begin{equation}
\xi^2 \frac{d^2 f}{d\xi ^2} + (a \xi^2 + b \xi) \frac{df}{d\xi} + (A\xi^2 + B\xi + C)f = 0 . \label{gCHGE}
\end{equation}
Here, we assume 
\begin{eqnarray}
a^2 - 4A \ne 0 . \label{assumCHGE}
\end{eqnarray}
Define $s$ and $\kappa $ as solutions of the quadratic equation
\begin{eqnarray}
&&\!\!\!\!\!\! s(s - 1) + b s + C = 0 , \\
&&\kappa ^2 + a \kappa + A = 0 . 
\end{eqnarray}
Because of assumption (\ref{assumCHGE}), $a + 2 \kappa \ne 0$. 

By the transformations of variables 
\begin{eqnarray}
\zeta = - (a + 2 \kappa) \xi , \\
f(\xi ) = \xi^s e^{\kappa \xi} \, v(\zeta ) , \label{gtransf}
\end{eqnarray}
Eq.\,(\ref{gCHGE}) transforms into the {\it canonical form} of the confluent hypergeometric equation: 
\begin{eqnarray}
\zeta \frac{d^2 v}{d\zeta^2} + (\gamma - \zeta)\frac{dv}{d\zeta} -\alpha v = 0 , \label{cCHGE}
\end{eqnarray}
where we write 
\begin{eqnarray}
\alpha \!&=&\! \frac{B + a s + b \kappa + 2 s \kappa }{a + 2 \kappa} , \\
\gamma \!&=&\! b + 2 s . 
\end{eqnarray}
If $\gamma $ is not an integer, two linearly independent solutions $v_1$ and $v_2$ of the confluent hypergeometric equation (\ref{cCHGE}) are written by means of the confluent hypergeometric function $F(\alpha , \gamma ; \zeta )$: 
\begin{eqnarray}
v_1 \!&=&\! F(\alpha , \gamma ; \zeta ) , \label{generalv1} \\
v_2 \!&=&\! \zeta ^{1 - \gamma } F(\alpha - \gamma +1 , 2 - \gamma ; \zeta ) . \label{generalv2}
\end{eqnarray}
The confluent hypergeometric function is defined by the confluent hypergeometric series: 
\begin{eqnarray}
F(\alpha , \gamma ; \zeta ) \!&=&\! \sum_{n=0}^{\infty} \frac{\alpha (\alpha + 1)\cdots (\alpha + n - 1)}{n!\, \gamma(\gamma + 1)\cdots (\gamma + n - 1)} \zeta^n ,  
\end{eqnarray}
and its analytic continuation. The term $F(\alpha, \gamma ; \zeta)$ is sometimes called Kummer's function. Substituting Eqs.\,(\ref{generalv1}) and (\ref{generalv2}) into Eq.\,(\ref{gtransf}), we obtain the general solution of Eq.\,(\ref{gCHGE}): 
\begin{eqnarray}
f(\xi ) \!&=&\! C_1 f_1 (\xi ) + C_2 f_2 (\xi ) , \\ 
f_1 (\xi ) \!&=&\! \xi^s e^{\kappa \xi } F\big( \alpha , \gamma ; - (a + 2 \kappa) \xi \big) , \\
f_2 (\xi ) \!&=&\! \xi^{s + 1 - \gamma } e^{\kappa \xi } F\big( \alpha - \gamma + 1 , 2 - \gamma ; - (a + 2\kappa ) \xi \big) .  
\end{eqnarray}
Apparently, we obtain $2 \times 2 \times 2 = 8$ solutions of Eq.\,(\ref{gCHGE}) since each $s$ and $\kappa $ has two values: 
\begin{eqnarray}
s_{\pm} \!&=&\! \frac{1 - b \pm i \sqrt{4C - (b - 1)^2 }}{2} , \\
\kappa_{\pm} \!&=&\! \frac{-a \pm i \sqrt{4A - a^2 }}{2} .  
\end{eqnarray}
However, when $\kappa $ is one fixed value, one finds the relations 
\begin{eqnarray}
\gamma_- \!&=&\! 2 - \gamma_+ , \label{relationg}\\
\alpha_- \!&=&\! \alpha_+ - \gamma_+ + 1 , \label{relationa}\\
s_- \!&=&\! s_+ + 1 - \gamma_+ , \label{relations} 
\end{eqnarray}
where each subscript sign represents that of $s_{\pm}$. Using these relations, 
\begin{eqnarray}
f_{2+} (\xi ) \!&=&\! \xi^{s_+ + 1 - \gamma_+ } e^{\kappa \xi } F\big( \alpha_+ - \gamma_+ + 1 , 2 - \gamma_+ ; - (a + 2\kappa ) \xi \big) \\
	&=&\! \xi^{s_-} e^{\kappa \xi } F\big( \alpha_- , \gamma_- ; - (a + 2 \kappa) \xi \big) \\
	&=&\! f_{1-} (\xi ) .  
\end{eqnarray}

Therefore, the two solutions $f _{2\pm} (\xi)$ and $f _{1\mp} (\xi )$ are equivalent.\footnote{Furthermore, the exchange $\kappa_+ $ to $\kappa_- $ in $F(\alpha , \gamma ; \zeta )$ is represented as Kummer's transformation
$$F(\gamma - \alpha , \gamma ; -\zeta ) = e^{-\zeta } F(\alpha , \gamma ; \zeta ) ,$$
so that $f_1 (\xi)$ and $f_2 (\xi)$ are invariant in the exchange. } Below, we write $f _{1+} (\xi )$ and $f _{2+} (\xi )$ simply as $f_1 (\xi )$ and $f_2 (\xi )$, respectively. 

The confluent hypergeometric function has the asymptotic expansion around $| \zeta | = \infty $~\cite{HMF1964}: 
\begin{eqnarray}
F(\alpha , \gamma ; \zeta ) \!&\sim&\! \frac{\Gamma (\gamma )}{\Gamma (\gamma - \alpha )} \sum_{n = 0} ^{\infty} \frac{\Gamma (n + \alpha ) \Gamma (n + \alpha - \gamma + 1)}{\Gamma (\alpha ) \Gamma (\alpha - \gamma + 1)} \frac{(-\zeta )^{- n - \alpha }}{n!} \nonumber \\
&& + \frac{\Gamma (\gamma )}{\Gamma (\alpha )} e^{\zeta } \sum_{n = 0} ^{\infty} \frac{\Gamma (n + \gamma - \alpha ) \Gamma (n + 1 - \alpha ) }{\Gamma (\gamma - \alpha ) \Gamma (1 - \alpha ) }  \frac{\zeta^{- n + \alpha - \gamma }}{n!} \\
\!&\sim&\! \Gamma (\gamma ) \bigg( \frac{(-\zeta )^{-\alpha }}{\Gamma (\gamma - \alpha )} + \frac{\zeta^{\alpha - \gamma }}{\Gamma (\alpha )} e^{\zeta } \bigg) + \mathcal{O}\left( | \zeta |^{-1} \right) .  
\end{eqnarray}
Therefore, $f_1 $ and $f_2 $ asymptotically behave around infinity as 
\begin{eqnarray}
f_1 (\xi ) \!&\sim&\! (a + 2\kappa )^{- \alpha_+ } \frac{\Gamma (\gamma_+ )}{\Gamma (\gamma_+ - \alpha_+ )} \xi ^{s_+ - \alpha_+ } e^{\kappa \xi } \nonumber \\
&&+ (-a - 2\kappa )^{\alpha_+ - \gamma_+ } \frac{\Gamma (\gamma_+ )}{\Gamma (\alpha_+ )} \xi ^{s_+ + \alpha_+ - \gamma_+ } e^{- (a + \kappa )\xi } , \\
f_2 (\xi ) \!&\sim&\! (a + 2\kappa )^{- \alpha_- } \frac{\Gamma (\gamma_- )}{\Gamma (\gamma_- - \alpha_- )} \xi ^{s_- - \alpha_- } e^{\kappa \xi } \nonumber \\
&&+ (- a - 2\kappa )^{\alpha_- - \gamma_- } \frac{\Gamma (\gamma_- )}{\Gamma (\alpha_- )} \xi ^{s_- + \alpha_- - \gamma_- } e^{-(a + \kappa )\xi } . 
\end{eqnarray}
From relations (\ref{relationg}), (\ref{relationa}), and (\ref{relations}), 
\begin{eqnarray}
s_+ - \alpha_+ \!\!\!&=&\!\!\! s_- - \alpha_- , \\
s_+ + \alpha_+ - \gamma_+ \!\!\!&=&\!\!\! s_- + \alpha_- - \gamma_- . 
\end{eqnarray}
Define the asymptotic coefficients $D_1 $ and $D_2 $ as 
\begin{equation}
C_1 f_1 (\xi ) + C_2 f_2 (\xi ) \ \sim \  D_1 \xi ^{s_+ - \alpha_+ } e^{\kappa \xi } + D_2 \xi ^{s_+ + \alpha_+ - \gamma_+ } e^{- (a + \kappa )\xi } . 
\end{equation}
We obtain the connection matrix $M$ between $(C_1 , C_2 )$ and $(D_1 , D_2 )$ as 
\begin{eqnarray}
&&\ \ \ \ \ \ \ \ \ \ \ \  \left(
	\begin{array}{l}
	D_1 \\
	D_2
	\end{array}
\right) 
= M 
\left(
	\begin{array}{l}
	C_1 \\
	C_2
	\end{array}
\right) , \\
M \!\!&=&\!\! \left(
	\begin{array}{cc}
	(a + 2\kappa )^{- \alpha_+ } \displaystyle \frac{\Gamma (\gamma_+ )}{\Gamma (\gamma_+ - \alpha_+ )} & (a + 2\kappa )^{- \alpha_- } \displaystyle \frac{\Gamma (\gamma_- )}{\Gamma (\gamma_- - \alpha_- )} \\
	(- a - 2\kappa )^{\alpha_+ - \gamma_+ } \displaystyle \frac{\Gamma (\gamma_+ )}{\Gamma (\alpha_+ )}  & (- a - 2\kappa )^{\alpha_- - \gamma_- } \displaystyle \frac{\Gamma (\gamma_- )}{\Gamma (\alpha_- )}
	\end{array}
\right) . 
\end{eqnarray}
Using Euler's reflection formula
\begin{eqnarray}
\Gamma(z) \Gamma(1 - z) = \frac{\pi }{\sin(\pi z)}
\end{eqnarray}
and 
\begin{eqnarray}
(-1)^z = \cos(\pi z) + i \sin(\pi z) , 
\end{eqnarray}
the determinant of $M$ is calculated as 
\begin{eqnarray}
\det{M} = \frac{\gamma_+ - 1}{a + 2 \kappa } . 
\end{eqnarray}
The inverse of $M$ is 
\begin{eqnarray}
M^{-1} = 
\left(
	\begin{array}{cc}
	(- a - 2\kappa )^{\alpha_+ } \displaystyle \frac{\Gamma (1 - \gamma_+ )}{\Gamma (\alpha_+ - \gamma_+ + 1)} & (a + 2\kappa )^{\gamma_+ - \alpha_+ } \displaystyle \frac{\Gamma (1 - \gamma_+ )}{\Gamma (1 - \alpha_+ )} \\
	(- a - 2\kappa )^{\alpha_- } \displaystyle \frac{\Gamma (1 - \gamma_- )}{\Gamma (\alpha_- - \gamma_- + 1)} & (a + 2\kappa )^{\gamma_- - \alpha_- } \displaystyle \frac{\Gamma (1 - \gamma_- )}{\Gamma (1 - \alpha_- )} 
	\end{array}
\right) . \label{Minverse}
\end{eqnarray}

\subsubsection{For critical modes}\label{secextcrit}
For critical modes, since $P_H = 0$, $b_{-1} = b_{-2} = 0$. The radial equation (\ref{DCHE}) reduces to 
\begin{eqnarray}
\xi^2 \frac{d^2 f}{d\xi ^2} + (a \xi^2 + b \xi ) \frac{df}{d\xi} + (A\xi^2 + B\xi + C)f = 0 , \label{extcrit}
\end{eqnarray}
where the parameters are explicitly written as 
\begin{eqnarray}
a \!&=&\! 0 , \\
b \!&=&\! 2 , \\
A \!&=&\! b_2 \, =\,  \eps^2 - \mu^2 , \\
B \!&=&\! b_1 \, =\,  2M(2\eps^2 - \mu^2) - 2\eps eQ , \\
C \!&=&\! b_0 \, =\,  (2M\eps - eQ )^2 - M^2 \mu^2 - \lam . 
\end{eqnarray}
As we saw in section \ref{secCHGE}, when $a^2 - 4A \ne 0$, i.e., 
\begin{eqnarray}
b_2 = \eps^2 - \mu^2 \ne 0 , 
\end{eqnarray}
one can transform Eq.\,(\ref{extcrit}) into the canonical form of the confluent hypergeometric equation. 

The quantities $s_{\pm}$ and $\kappa $ are 
\begin{eqnarray}
s_{\pm} \!&=&\! \frac{-1 \pm i \sqrt{4b_0 - 1}}{2} , \\
\kappa \!&=&\! i \sqrt{\eps^2 - \mu^2 }  
\end{eqnarray}
and $\alpha_{\pm }$ and $\gamma_{\pm}$ are 
\begin{eqnarray}
\alpha_{\pm } \!&=&\! \frac{b_1 }{2\kappa } + 1 + s_{\pm } , \\
\gamma_{\pm } \!&=&\! 2 + 2s_{\pm } . 
\end{eqnarray}
Two solutions of Eq.\,(\ref{extcrit}), $f_1 (\xi )$ and $f_2 (\xi )$, are obtained:  

\begin{eqnarray}
f_1 (\xi ) \!&=&\! \xi^{s_+ } e^{\kappa \xi } F( \alpha_+ , \gamma_+ ; - 2 \kappa \xi ) \label{foutPH=0} \\
	 \!&\sim&\! ( 2\kappa )^{- \alpha_+ } \frac{\Gamma (\gamma_+ )}{\Gamma (\gamma_+ - \alpha_+ )} \xi ^{s_+ - \alpha_+ } e^{\kappa \xi } \nonumber \\
	 &&+ ( - 2\kappa )^{\alpha_+ - \gamma_+ } \frac{\Gamma (\gamma_+ )}{\Gamma (\alpha_+ )} \xi ^{s_+ + \alpha_+ - \gamma_+ } e^{- \kappa \xi } , \\
f_2 (\xi ) \!&=&\! f_{1-} (\xi )\\
	\!&=&\! \xi^{s_- } e^{\kappa \xi } F( \alpha_- , \gamma_- ; - 2 \kappa \xi ) \label{finPH=0} \\
	\!&\sim&\! ( 2\kappa )^{- \alpha_- } \frac{\Gamma (\gamma_- )}{\Gamma (\gamma_- - \alpha_- )} \xi ^{s_- - \alpha_- } e^{\kappa \xi } \nonumber \\
	&&+ ( - 2\kappa )^{\alpha_- - \gamma_- } \frac{\Gamma (\gamma_- )}{\Gamma (\alpha_- )} \xi ^{s_- + \alpha_- - \gamma_- } e^{- \kappa \xi } .  \label{fi asym extcrit}
\end{eqnarray}

For critical modes $P_H = 0$, the definition of ingoing(outgoing) modes at the event horizon depends on whether $P_H = + 0$ or $- 0$.

In the extremal case, 
\begin{eqnarray}
\pi_r ^2 = \frac{1}{\xi^2 } \left( b_2 \xi^2 + b_1 \xi + b_0 + \frac{b_{- 1} }{\xi } + \frac{b_{- 2} }{\xi^2 } \right) . 
\end{eqnarray}
For critical particles $P_H = 0$, since $b_{- 1} = b_{- 2} = 0$, 
\begin{eqnarray}
\pi_r = \frac{\sig_r }{\xi } \sqrt{b_0 + b_1 \xi + b_2 \xi^2 } . 
\end{eqnarray}
From Eqs.\,(\ref{foutPH=0}) and Eq.\,(\ref{finPH=0}), the logarithmic derivatives of $f_1 (\xi )$ and $f_2 (\xi )$ are 
\begin{eqnarray}
\del_r \ln f_1 (\xi ) \!&=&\! \frac{s_+ }{\xi } + \kappa - 2\kappa \frac{\alpha_+ }{\gamma_+ } \frac{F(\alpha_+ + 1 , \gamma_+ + 1 ; - 2\kappa \xi )}{F(\alpha_+ , \gamma_+ ; - 2\kappa \xi )} , \label{logderivf1crit} \\
\del_r \ln f_2 (\xi ) \!&=&\! \frac{s_- }{\xi } + \kappa - 2\kappa \frac{\alpha_- }{\gamma_- } \frac{F(\alpha_- + 1 , \gamma_- + 1 ; - 2\kappa \xi )}{F(\alpha_- , \gamma_- ; - 2\kappa \xi )} . \label{logderivf2crit}
\end{eqnarray}
Writing $\hbar$ and $P_H = \pm 0$ explicitly, 
\begin{eqnarray}
- i \hbar P_H \del_r \ln f_1 (\xi ) \!&=&\! P_H \left( \sqrt{\frac{b_0 }{\xi^2 } - \frac{\hbar^2 }{4\xi^2 } } + \frac{i \hbar }{2\xi } \right)  + \mathcal{O}(\xi^0 ) \\
	&=&\! P_H \left( \sqrt{\pi_r ^2 - \frac{\hbar^2 }{4\xi^2 } - \frac{b_1 }{\xi } - b_2 } + \frac{i \hbar }{2\xi } \right) + \mathcal{O}(\xi^0 ) , \\
- i \hbar P_H \del_r \ln f_2 (\xi ) \!&=&\! - P_H \left( \sqrt{\frac{b_0 }{\xi^2 } - \frac{\hbar^2 }{4\xi^2 } } - \frac{i \hbar }{2\xi } \right) + \mathcal{O}(\xi^0 ) \\
	&=&\! - P_H \left( \sqrt{\pi_r ^2 - \frac{\hbar^2 }{4\xi^2 } - \frac{b_1 }{\xi } - b_2 } - \frac{i \hbar }{2\xi } \right) + \mathcal{O}(\xi^0 ) . 
\end{eqnarray}
It follows that when $P_H = + 0$, $f_1 (\xi )$ is pure outgoing and $f_2 (\xi )$ is pure ingoing at the horizon. However, when $P_H = - 0$ and $b_0 > \frac{\hbar^2 }{4}$, then $f_1 (\xi )$ is pure ingoing and $f_2 (\xi )$ is pure outgoing, and when $P_H = - 0$ and $b_0 \le \frac{\hbar^2 }{4}$, then $f_1 (\xi )$ is tunneling in the outgoing direction and $f_2 (\xi )$ is tunneling in the ingoing direction at the horizon. We define $f^{out} (\xi )$ and $f^{in} (\xi )$ by 
\begin{eqnarray}
f^{out} (\xi ) \!&=&\! f_1 (\xi ), \ \ f^{in} (\xi ) \,=\, f_2 (\xi ) \ \ \ \ \ \ \ \ \ \left({\rm when\ \ }P_H = + 0 \ \ {\rm or}\ \ b_0 \le \frac{\hbar^2 }{4} \right) , \\
f^{out} (\xi ) \!&=&\! f_2 (\xi ), \ \  f^{in} (\xi ) \,=\, f_1 (\xi ) \ \ \ \ \ \ \ \ \ \left({\rm when\ }P_H = - 0 \ \ {\rm and}\ \ b_0 > \frac{\hbar^2 }{4}  \right) . 
\end{eqnarray} 

When $\eps^2 > \mu^2 $, namely $\kappa $ is pure imaginary, $N f_i (\xi )$ ($i = 1,\,2$) denote solutions that contain pure ingoing modes at the event horizon $\xi = 0$ and which are normalized up to scale such that the amplitudes of the ingoing modes at $\xi = \xi_0 \gg M$ are $\sim C_0 ^{in} $. Their explicit forms and asymptotic behaviors are 
\begin{eqnarray}
N f_1 (\xi ) \!&=&\! C_0 ^{in} (-2\kappa )^{\gamma_+ - \alpha_+ } \frac{\Gamma(\alpha_+ )}{\Gamma(\gamma_+ )} \xi_0 ^{-s_+ - \alpha_+ + \gamma_+ } \xi^{s_+ } e^{\kappa \xi } F\big( \alpha_+ , \gamma_+ ; - 2 \kappa \xi \big) \label{foextcrit} \\
	&\sim&\! C_0 ^{in} (-1)^{\gamma_+ - \alpha_+ } (2\kappa )^{\gamma_+ - 2 \alpha_+ } \frac{\Gamma(\alpha_+ )}{\Gamma(\gamma_+ - \alpha_+ ) } \xi_0 ^{- \frac{B}{2\kappa } + 1} \xi ^{- \frac{B}{2\kappa } - 1} e^{\kappa \xi } \nonumber \\
	&& + \, C_0 ^{in} \left( \frac{\xi }{\xi_0 } \right) ^{\frac{B}{2\kappa } - 1} e^{- \kappa \xi } , \\
N f_2 (\xi ) \!&=&\! C_0 ^{in} (-2\kappa )^{\gamma_- - \alpha_- } \frac{\Gamma(\alpha_- )}{\Gamma(\gamma_- )} \xi_0 ^{-s_- - \alpha_- + \gamma_- } \xi^{s_{-} } e^{\kappa \xi } F\big( \alpha_- , \gamma_- ; - 2 \kappa \xi \big) \label{fiextcrit} \\
	&\sim&\! C_0 ^{in} (-1)^{\gamma_- - \alpha_- } (2\kappa )^{\gamma_- - 2 \alpha_- } \frac{\Gamma(\alpha_- )}{\Gamma(\gamma_- - \alpha_- ) } \xi_0 ^{- \frac{B}{2\kappa } + 1} \xi ^{- \frac{B}{2\kappa } - 1} e^{\kappa \xi } \nonumber \\
	&& + \, C_0 ^{in} \left( \frac{\xi }{\xi_0 } \right) ^{\frac{B}{2\kappa } - 1} e^{- \kappa \xi } . 
\end{eqnarray}

Note that a series of interesting solutions can be found. If $\alpha_- $ is a nonpositive integer $- n$, i.e.,\footnote{Condition (\ref{condbound}) is equivalent to the extremal limit of the characteristic resonance condition in~\cite{Hod2014b}.}  
\begin{eqnarray}
 &\mu^2 > \eps^2 , \ \ \ \ \ \  4b_0 - 1 \le 0 , & \nonumber \\
 &2 \left[ M(2 \eps^2 - \mu^2 ) - \eps e Q \right] = \sqrt{\mu^2 - \eps^2 } \left( \sqrt{1 - 4 b_0 } + 2 n + 1 \right) ,& \label{condbound}  
\end{eqnarray}
where $n = 0,\,1,\,2,\dots$, then, the confluent hypergeometric series 
\begin{eqnarray}
F( - n , \gamma_- ; -2\kappa \xi ) = \sum_{m=0}^{ \infty } \frac{- n (- n + 1)\dots (- n + m - 1)}{m!\, \gamma_- (\gamma_- + 1)\dots (\gamma_- + m - 1)} (-2\kappa \xi )^m 
\end{eqnarray}
stops at $m = n$ so that $F( - n , \gamma_- ; -2\kappa \xi )$ is a polynomial of degree $n$. Therefore, each 
\begin{eqnarray}
f_2 (\xi ) = \xi ^{\frac{-1 + \sqrt{1 - 4b_0 }}{2}} e^{-\xi \sqrt{\mu^2 - \eps^2 } } F( - n , \gamma_- ; -2\kappa \xi )
\end{eqnarray}
decays exponentially as $\xi \rightarrow \infty $, and describes a bound state.

\subsubsection{For marginal modes with specific charges}\label{secmarginal}
For marginal modes that satisfy
\begin{eqnarray}
\eps^2 = \mu^2  \ \ \ \ \ \ {\rm and}\ \ \ \ \ \  \eps (\eps M - e Q ) = 0 , \label{extmarginal}
\end{eqnarray}
Eq.\,(\ref{DCHE}) reduces to 
\begin{eqnarray}
\xi^2 \frac{d^2 f}{d\xi^2} + 2\xi \frac{df}{d\xi } + \left( b_0 + \frac{b_{-1} }{\xi } + \frac{b_{-2}}{\xi^2 } \right) f = 0 . \label{preCHGE}
\end{eqnarray}
By the transformation of variables 
\begin{eqnarray}
\xi = \frac{1}{z} , 
\end{eqnarray}
Eq.\,(\ref{preCHGE}) transforms into 
\begin{eqnarray}
z^2 \frac{d^2 f}{dz^2 } + \left( az^2 + bz \right) \frac{df}{dz} + \left( A z^2 + B z + C \right) f = 0 , 
\end{eqnarray}
where the parameters are explicitly written as 
\begin{eqnarray}
a \!&=&\! 0 , \\
b \!&=&\! 0 , \\
A \!&=&\! b_{-2} \,=\, P_H ^2 , \\ 
B \!&=&\! b_{-1} \,=\, 2 P^{\prime }_H P_H , \\ 
C \!&=&\! b_0 \ =\  2a \eps (a \eps - m ) + Q^2 e^2 - M^2 \eps^2 - \lam . 
\end{eqnarray}
When $a^2 - 4A \ne 0$, i.e.,  
\begin{eqnarray}
b_{- 2} = P_H ^2 \ne 0 , 
\end{eqnarray}
the quantities $s_{\pm}$ and $\kappa $ are 
\begin{eqnarray}
s_{\pm} \!&=&\! \frac{1 \pm i \sqrt{4b_0 - 1}}{2} , \\
\kappa \!&=&\!  i P_H 
\end{eqnarray}
and $\alpha_{\pm} $ and $\gamma_{\pm} $ are 
\begin{eqnarray}
\alpha_{\pm} \!&=&\! - i P^{\prime }_H  + s_{\pm } , \\ 
\gamma _{\pm } \!&=&\! 2 s_{\pm } . 
\end{eqnarray}

Using the formula (\ref{Minverse}), solutions $f^{out} (\xi )$ and $f^{in}(\xi )$ that contain pure outgoing and ingoing modes at the event horizon $\xi = 0$, respectively, are identified up to scale: 
\begin{eqnarray}
f^{out} (\xi ) \!&=&\! (2\kappa )^{\gamma_+ - \alpha_+ } \frac{\Gamma(1 - \gamma_+ )}{\Gamma(1 - \alpha_+ )} \xi^{- s_+ } e^{\frac{\kappa }{\xi }} F\Big( \alpha_+ , \gamma_+ ; - \frac{2\kappa }{\xi }  \Big) \nonumber \\ 
	&& \ \ \ \ \ \ + (2\kappa )^{\gamma_- - \alpha_- } \frac{\Gamma(1 - \gamma_- )}{\Gamma(1 - \alpha_- )} \xi^{ - s_- } e^{\frac{\kappa }{\xi }} F\Big( \alpha_- , \gamma_- ; - \frac{2\kappa }{\xi }  \Big) \label{fopure} \\
		&\sim& \xi^{ i P^{\prime }_H } e^{\frac{- \kappa }{\xi }} \ \ \ \ \ \ \ \ \ \ \ \ \text{(near horizon)} , \\
f^{in} (\xi ) \!&=&\! (- 2\kappa )^{ \alpha_+ } \frac{\Gamma(1 - \gamma_+ )}{\Gamma(\alpha_+ - \gamma_+ + 1 )} \xi^{- s_+ } e^{\frac{\kappa }{\xi }} F\Big( \alpha_+ , \gamma_+ ; - \frac{2\kappa }{\xi }  \Big) \nonumber \\ 
	&& \ \ \ \ \ \ + (- 2\kappa )^{ \alpha_- } \frac{\Gamma(1 - \gamma_- )}{\Gamma( \alpha_- - \gamma_- + 1 )} \xi^{ - s_- } e^{\frac{\kappa }{\xi }} F\Big( \alpha_- , \gamma_- ; - \frac{2\kappa }{\xi }  \Big) \label{fipure} \\
		&\sim& \xi^{- i P^{\prime }_H } e^{\frac{\kappa }{\xi }} \ \ \ \ \ \ \ \ \ \ \ \ \ \, \text{(near horizon)} . 
\end{eqnarray}
Note that $f^{in}(\xi )$ relates to Tricomi's function. Using Tricomi's function, 
\begin{eqnarray}
U(a, b, z) = \frac{\Gamma(1 - b)}{\Gamma(1 + a - b)} F(a, b; z) + \frac{\Gamma(b - 1)}{\Gamma(a)} z^{1 - b} F(1 + a - b, 2 - b; z) , \label{Tricomi}
\end{eqnarray}
$f^{in} (\xi )$ is written as 
\begin{eqnarray}
f^{in} (\xi ) = (-2\kappa )^{\alpha_- } \xi^{- s_- } e^{\frac{\kappa }{\xi }} U\Big( \alpha_- , \gamma_- , - \frac{2\kappa }{\xi } \Big) . 
\end{eqnarray}

The first terms of the right-hand side of Eqs.\,(\ref{fopure}) and (\ref{fipure}) represent ingoing modes at a sufficiently distant point.  We adopt the normalization procedure in which $N f^{in} (\xi )$ contains a pure ingoing mode at $\xi = 0$ and whose amplitude of ingoing mode at $\xi = \xi_0 $ is $\sim C_0^{in}$. The normalization factor $N$ is 
\begin{eqnarray}
N = C_0^{in} (- 2\kappa )^{- \alpha_+ } \frac{\Gamma(\alpha_+ - \gamma_+ + 1)}{\Gamma(1 - \gamma_+ )} \xi_0^{ s_+ } .  
\end{eqnarray}

Tricomi's function (\ref{Tricomi}) is defined even when $a$ is a nonpositive integer by the limiting procedure. For $b = n + 1$, $n = 0,\, 1, \, 2, \dots $ and $a \ne 0, \, -1, \, -2, \dots $, Tricomi's function $U(a, n + 1 , z)$ is~\cite{HMF1964}
\begin{eqnarray}
U(a , n + 1 , z) \!&=&\! \frac{(- 1)^{n + 1}}{n! \G(a - n)} \sum_{k = 0} ^{\infty } \frac{(a)_k }{(n + 1)_k k! } z^k [\ln z + \psi (a + k) - \psi (1 + k) - \psi (n + k + 1) ] \nonumber \\
&& + \frac{1}{\G(a)} \sum_{k = 1} ^n \frac{(k - 1)! (1 - a + k)_{n - k}}{(n - k)! } z^{- k} . 
\end{eqnarray}
For example, when $\gamma_- = 1$ (or $4b_0 = 1$), then 
\begin{eqnarray}
f^{in} (\xi ) = - \frac{(-2\kappa )^{\alpha_- } }{\G(\alpha_- )} \xi^{- \frac{1}{2}} e^{\frac{\kappa }{\xi }} \sum_{k = 0} ^{\infty } \frac{(\alpha_- )_k }{(k !)^2 } \bigg( \frac{- 2\kappa }{\xi } \bigg)^k [ \ln z + \psi (\alpha_- + k) - 2\psi (1 + k) ] . 
\end{eqnarray}
This solution behaves $\sim \xi^{- \frac{1}{2}} $ in the far region.

\subsubsection{For critical and marginal modes}\label{seccritmarginal}
The radial equation for the critical modes (\ref{extcrit}) can reduce to a simpler equation under additional conditions. In sections \ref{seccritmarginal} and \ref{seccritmarginalspec}, for simplicity, we assume $P_H = + 0$. 

When $P_H = + 0$ and $\eps^2 = \mu^2 $, i.e., a critical marginal mode, Eq.\,(\ref{extcrit}) reduces to 
\begin{eqnarray}
\xi^2 \frac{d^2 f}{d\xi^2 } + 2\xi \frac{df}{d\xi } + \left( b_1 \xi + b_0 \right) f = 0 ,  
\end{eqnarray}
where
\begin{eqnarray}
b_1 \!&=&\! 2\eps (M\eps - eQ) , \\ 
b_0 \!&=&\! (3M\eps - eQ)(M\eps - eQ) - \lam . 
\end{eqnarray}
When $b_1 = 2\eps (M\eps - Q e) \neq 0$, by the transformations of variables
\begin{eqnarray}
\xi = \frac{y^2 }{4b_1 }  \ \ \ \ \ \ {\rm and} \ \ \ \ \ \ f(\xi ) = \frac{2 \sqrt{b_1 } }{ y } \tilde{f}(y) , 
\end{eqnarray}
Eq.\,(\ref{extcrit}) transforms to 
\begin{eqnarray}
\frac{d^2 \tilde{f} }{dy^2 } + \frac{1}{y} \frac{d\tilde{f}}{dy} + \left( 1 - \frac{1 - 4 b_0 }{y^2 } \right) \tilde{f} = 0 . \label{Bessel}
\end{eqnarray}
Equation (\ref{Bessel}) is the {\it Bessel equation}, which has two fundamental solutions:\footnote{When $\nu $ is an integer, $J_{\nu} (y)$ and $J_{-\nu } (y)$ are not linearly independent.} 
\begin{eqnarray}
\tilde{f}_{\pm } (y) \!&=&\! J_{\pm \nu } (y) \\
	&=&\! \left( \frac{y}{2} \right)^{\pm \nu } \sum_{m=0}^{\infty } \frac{(- 1)^m }{m!\, \Gamma(m \pm \nu + 1)} \left( \frac{y}{2} \right)^{2m} ,  \label{defBessel}
\end{eqnarray}
where $\nu = i \sqrt{4b_0 - 1 }$. The function $J_{\nu }(y)$ is called the {\it Bessel function}. The radius of convergence of the power series (\ref{defBessel}) is $\infty $, so the Bessel function is defined in the entire space. 

Their asymptotic behaviors are~\cite{HMF1964} 
\begin{eqnarray}
\tilde{f}_{\pm} (y) \!&\sim&\! \frac{1}{\Gamma(\pm \nu + 1)} \left( \frac{y}{2} \right)^{\pm \nu } \ \ \ \ \ \ \ \ \ \ \ \ \ \ \ \ \ \ \ \ \  (\ y \sim 0 \ ) , \label{asympextp0Bessel} \\
\tilde{f}_{\pm} (y) \!&\sim&\! \sqrt{\frac{2}{\pi y}} \cos\!\left( y \mp \frac{ \nu \pi }{2} - \frac{\pi }{4} \right) \  \ \ \ \ \ \ \ \ \ \ \ \ \ (\ y \sim \infty ) \\
	&=&\! \sqrt{\frac{1}{2\pi y}} \exp \left[ i \left( y \mp \frac{ \nu \pi }{2} - \frac{\pi }{4} \right) \right] +  \sqrt{\frac{1}{2\pi y}} \exp \left[ - i \left( y \mp \frac{ \nu \pi }{2} - \frac{\pi }{4} \right) \right] . 
\end{eqnarray}
The normalized solution, which is pure ingoing at the horizon $\xi = 0$ and whose ingoing amplitude at a distant point $\xi = \xi_0 \gg M$, is $\sim C_0^{in} $ is obtained: 
\begin{eqnarray}
Nf^{in} (\xi ) \!&=&\! C_0^{in} 2 \sqrt{\pi} \xi_0 \left( \frac{b_1 }{\xi_0 } \right)^{\frac{1}{4}} \xi^{- \frac{1}{2}} J_{- \nu } \big( 2\sqrt{b_1 \xi } \big) \\
	&\sim&\! \left\{ 
	\begin{array}{l}
	C_0^{in} 2 \sqrt{\pi } \xi_0 \left( \frac{b_1 }{\xi_0 } \right)^{\frac{1}{4}} \frac{b_1 ^{- \frac{\nu }{2}}}{\Gamma(- \nu + 1)} \xi^{- \frac{1 + \nu }{2} } \ \ \ \ \ \ \ \ \ \ (\ \xi \sim 0 \ ) , \\
	2 C_0^{in} \left( \frac{\xi_0 }{\xi } \right)^{\frac{3}{4}} \cos\!\left( y + \frac{ \nu \pi }{2} - \frac{\pi }{4} \right) . \ \ \ \ \ \ \ \ \ \ \ \  (\ \xi \sim \infty ) . 
	\end{array}
	\right.
\end{eqnarray}

\subsubsection{For critical and marginal modes with specific charges}\label{seccritmarginalspec}
 When $P_H = + 0 $, $\eps^2 = \mu^2 $, and $\eps ( M \eps - Q e) = 0$, Eq.\,(\ref{extcrit}) reduces to 
\begin{eqnarray}
\xi^2 \frac{d^2 f}{d\xi^2 } + 2 \xi \frac{df}{d\xi } + b_0 f = 0 , \label{Euler}
\end{eqnarray}
which is an Euler equation. Here, 
\begin{eqnarray}
b_0 =  \left \{
\begin{array}{l}
Q^2 e^2 - \lam \ \ \ \ \ \ \ (\text{when }\  \eps = 0 ) , \\
- \lam \ \ \ \ \ \ \ \ \ \ \ \ \ \ \ (\text{when }\, M\eps - Q e = 0 ) . 
\end{array}
\right.
\end{eqnarray}

The general solution is  
\begin{eqnarray}
f(\xi) = C_+ \xi^{ n_+ } + C_- \xi ^{ n_- } , \label{solutionEuler}
\end{eqnarray}
where $n_{\pm } = \displaystyle\frac{-1 \pm i \sqrt{ 4b_0 - 1}}{2}$. The solution (\ref{solutionEuler}) is a superposition of two modes outgoing(the first term) and ingoing(the second term) in the entire space. The ingoing normalized solution $Nf^{in}(\xi )$, whose amplitude at $\xi = \xi_0 $ is $C_0^{in}$, is 
\begin{eqnarray}
Nf^{in} (\xi ) = C_0^{in} \left( \frac{\xi }{\xi_0 } \right)^{n_- } . 
\end{eqnarray}


\section{The intensity of collision of two scalar field excitations}\label{trace}
Our aim is to calculate the field-theoretical counterpart of the collision energy of two particles in the center-of-mass frame. However, the meaning of such a quantity, i.e., the ``collision energy'' of two excitations of fields, is somewhat unclear. 

For example, if only the consistency in the classical limit is respected, one might replace $\pi_{i a}$ by $\del_a S_i$ in Eq.\,(\ref{defEcm}). But one can also replace it by $\del_a \bar{S}_i $, or $\pi_{1a}$ and $\pi_{2a}$ by $\del_a S_1 $ and $\del_a \bar{S}_2$ respectively, and so on. Furthermore, adding any quantities that go to $0$ in the classical limit does not affect the consistency. Therefore, these formal replacements without reasonable physical meanings tell us nothing about the quantum effect, and are outside our interest. 

In this paper, we use the scalar norm of the sum of covariant 4-gradients of two fields, 
\begin{eqnarray}
E^2 = \hbar^2 g^{ab} \left( \bar{D}_a \bar{\Psi}_1 + \bar{D}_a \bar{\Psi}_2 \right) \left( D_b \Psi_1 + D_b \Psi_2 \right) , 
\end{eqnarray} 
as an intensity of the ``collision'', where $D_a $ is the covariant derivative $D_a \Psi_i = \del_a \Psi_i - \frac{i}{\hbar} e_i A_a \Psi_i $. The quantity $E^2 $ has several desirable properties. First, this is a locally defined scalar, hence independent of the coordinate.  

Second, when $\Psi_i = \exp\left(\frac{i}{\hbar } S_i \right) $, 
\begin{eqnarray}
E^2 \!\!\!&=&\!\!\! \left( \del_a \bar{S}_1 - e_1 A_a \right) \left( \del^a S_1 - e_1 A^a \right) \bar{\Psi}_1 \Psi_1 \nonumber \\
 &&\!\!\!\!\! + \left( \del_a \bar{S}_2 - e_2 A_a \right) \left( \del^a S_2 - e_2 A^a \right) \bar{\Psi}_2 \Psi_2 \nonumber \\
 &&\!\!\!\!\! + \left( \del_a \bar{S}_1 - e_1 A_a \right) \left( \del^a S_2 - e_2 A^a \right) \bar{\Psi}_1 \Psi_2 \nonumber \\
 &&\!\!\!\!\! + \left( \del_a \bar{S}_2 - e_2 A_a \right) \left( \del^a S_1 - e_1 A^a \right) \bar{\Psi}_2 \Psi_1 , 
\end{eqnarray}
so, under the condition $\Psi_1 = \Psi_2 = 1$, $E^2 $ formally transforms into $E_{\rm cm} ^2 $ in the classical limit $\hbar \rightarrow 0$.  

Third, when $\mu_1 = \mu_2 = \mu $, $e_1 = e_2 = e$, $E^2$ is related to the stress tensor of $\Psi = \Psi_1 + \Psi_2 $. The stress tensor $T_{ab}$ and its trace $T$ are written as  
\begin{eqnarray}
T_{ab} \!&=&\! \hbar^2 \bar{D}_a \bar{\Psi} D_b \Psi - \frac{\hbar^2 }{2} g_{ab} g^{cd} \bar{D}_c \bar{\Psi} D_d \Psi - \frac{1}{2} \mu^2 g_{ab} \bar{\Psi} \Psi , \\
T \!&=&\! g^{ab} T_{ab} \\
\!&=&\! - \hbar^2 g^{ab} \bar{D}_a \bar{\Psi} D_b \Psi - 2 \mu^2 \bar{\Psi} \Psi \\
 &=&\! - E^2 - 2 \mu^2 \bar{\Psi} \Psi . 
\end{eqnarray}
Therefore, we arrive at calculations of the trace of the stress tensor. The physical meaning of the stress tensor is clear: the source of the back-reaction to the gravitational field. It is possible to say that we define the intensity of a ``collision'' of field excitations by an intensity of the back-reaction to the gravitational field. 

To calculate $E^2 $ at the event horizon, we must obtain the values of $g^{ab} (\del_a \bar{S}_1 - e_1 A_a )( \del_b S_2 - e_2 A_b ) $ and $\bar{\Psi }_1 \Psi_2 $ there. The former can be obtained only by asymptotic behaviors of the local solutions around the event horizon\footnote{Because this calculation requires the next-to-leading order of the asymptotic expansion of $f(r)$, approximate solutions may derive wrong results.}. In sections \ref{secIntensityNonext} and \ref{secIntensityext}, we calculate that in general situations. 

On the other hand, the latter quantity $\bar{\Psi }_1 \Psi_2$ requires global information about the solutions. Since we have obtained several global solutions of the radial part of the Klein-Gordon equation in some specific cases, let us evaluate that in each case.

\subsection{Amplitude of the radial part of the scalar field}
Summarizing the previous section, the amplitudes of the appropriately normalized solutions $Nf^{in} $ at the event horizon are as follows: 
\begin{enumerate}
	\item When $r_H \ne r_C $, $\eps^2 = \mu^2 $, and $\eps (M \eps - Q e) = 0$, 
		\begin{eqnarray}
		&& C_0^{in} \frac{\Gamma(a) \Gamma(c - b)}{\Gamma(c) \Gamma(a - b)} (- x_0 )^{- A }, \\
		&& {\rm where}\ \ - A = \frac{1}{2} + \frac{i}{2} \sqrt{4\big[ 2\eps P_H + (2M\eps - Q e)^2 - r_H ^2 \mu^2 - \lam \big] - 1} . \nonumber 
		\end{eqnarray}
	\item When $r_H = r_C $, $\eps^2 \ne \mu^2 $, and $P_H = + 0$,  
		\begin{eqnarray}
		C_0 ^{in} (-2\kappa )^{\gamma_- - \alpha_- } \frac{\Gamma(\alpha_- )}{\Gamma(\gamma_- )} \xi_0 ^{- s_- - \alpha_- + \gamma_- } \xi^{s_{-} } ,  \ \ \ \ {\rm where}\ \ \ s_- = \frac{-1 - i \sqrt{4 b_0 - 1}}{2} . 
		\end{eqnarray}
	\item When $r_H = r_C $, $\eps^2 = \mu^2 $, $P_H \ne 0$, and $\eps (M \eps - Q e ) = 0$, 
		\begin{eqnarray}
		 C_0^{in} (- 2\kappa )^{- \alpha_+ } \frac{\Gamma(\alpha_+ - \gamma_+ + 1)}{\Gamma(1 - \gamma_+ )} \xi_0^{ s_+ } ,  \ \ \ \ {\rm where}\ \ \ s_+ = \frac{1 + i \sqrt{4 b_0 - 1}}{2} . \label{case3}
		\end{eqnarray}
	\item When $r_H = r_C $, $\eps^2 = \mu^2$, $P_H = + 0$, and $\eps (M\eps - Q e) \ne 0$,  
		\begin{eqnarray}
		C_0^{in} 2 \sqrt{\pi } \xi_0 \left( \frac{b_1 }{\xi_0 } \right)^{\frac{1}{4}} \frac{b_1 ^{- \frac{\nu }{2}}}{\Gamma( - \nu + 1)} \xi^{- \frac{1 + \nu }{2} }, \ \ \ \ {\rm where}\ \ \  - \frac{1 + \nu }{2} = \frac{-1 - i \sqrt{4 b_0 - 1}}{2} . 
		\end{eqnarray}
	\item When $r_H = r_C $, $\eps^2 = \mu^2 $, $P_H = + 0$, and $\eps (M \eps - Q e ) = 0$, 
		\begin{eqnarray}
		 C_0^{in} \left( \frac{\xi }{\xi_0 } \right)^{n_- } , \ \ \ \ \ \ \ \ \ \ \ \ \ \ \ {\rm where}\ \ \ n_- = \frac{-1 - i \sqrt{4 b_0 -1}}{2} . \label{case5}
		\end{eqnarray}
\end{enumerate}
Since gamma functions have no zeros but poles at nonpositive integers, in each case the amplitude at the horizon can be $0$. The results are in Table \ref{zeros}. In each case, unless these conditions are satisfied, the amplitude of the ingoing mode at the horizon does not vanish. Similarly, the divergences of the amplitudes are also given. 
\begin{table}[!h]
\caption{Zeros and divergences of the amplitude at the horizon.}
\label{zeros}
\centering
\begin{tabular}{l|c|c}
\hline
\multicolumn{1}{c}{Conditions} & \multicolumn{1}{|c|}{Zeros} & \multicolumn{1}{c}{Divergences} \\ 
\hline \hline
$r_H \ne r_C $, \ $\eps^2 = \mu^2 $, \ \ \ \ \ \ \ \ \ \ \ \ \ \ \ \,$\eps (M\eps - Q e) = 0$ & $a = b$ & No divergences \\
\hline
$r_H = r_C $, \ $\eps^2 \ne \mu^2 $, \ $P_H = + 0$ & $b_0 < 0$ & Eq.\,(\ref{condbound}) and $b_0 > 0 $ \\
\hline
$r_H = r_C $, \ $\eps^2 = \mu^2 $, \ $P_H \ne \pm 0$, \ $\eps (M\eps - Q e) = 0$ & $4b_0 = 1$ & No divergences \\
\hline
$r_H = r_C $, \ $\eps^2 = \mu^2$, \ $P_H = + 0$, \ $\eps (M\eps - Q e) \ne 0$ & $b_0 < 0$ & $b_0 > 0$ \\
\hline
$r_H = r_C $, \ $\eps^2 = \mu^2 $, \ $P_H = + 0$, \ $\eps (M \eps - Q e) = 0$ & $b_0 < 0$ & $b_0 > 0$ \\
\hline
\end{tabular}
\end{table}

In the second, fourth and fifth cases when $b_0 > 0$, the amplitudes at the horizon diverge as $\sim \xi ^{\frac{-1 - i \sqrt{4 b_0 - 1}}{2}} $. These divergences come from the fact that the turning point of a critical particle ($P_H = 0$) coincides with the horizon of the extremal black hole in the particle picture. At the turning point, the critical particle spends an infinitely long time and the probability that it is observed there diverges. On the other hand, when $b_0 < 0$, the amplitudes at the horizon are $0$. Because the condition $b_0 < 0$ corresponds to that of forbidden motions at the horizon and the turning points move to outside in the particle picture, the amplitudes decay in the ingoing direction. The proper distance between the horizon and an external point $\xi > 0$ is infinite in the extremal case. That is why in the condition $b_0 < 0$, the amplitudes at the horizon are $0$. 

When $a = b$ in the first case and $4b_0 = 1$ in the third case, the amplitudes at the horizon are also $0$. In these situations, as seen below Eq.\,(\ref{f1a=b}), the outgoing and ingoing modes at infinity are degenerate so that our normalization procedure becomes invalid. 

Note that these divergences do not always mean those of physical quantities. In practice, a field excitation forms a wave packet, so it is necessary to integrate the product of the physical quantity and a suitable wave function over the parameter.   

\ 

In the following subsections, we calculate $g^{ab} (\del_a \bar{S}_1 - e_1 A_a )( \del_b S_2 - e_2 A_b )$ from local solutions. For simplicity, we define 
\begin{eqnarray}
I_{ij} = 2 \bigg( \frac{P_i P_j }{\Delta} - \hbar^2 \Delta \del_r \ln \bar{f}_i \del_r \ln f_j \bigg) 
\end{eqnarray}
so that 
\begin{eqnarray}
&&g^{ab} (\del_a \bar{S}_i - e_i A_a )( \del_b S_j - e_2 A_j ) \nonumber \\
\!&=&\! -\frac{1}{\rho^2} \bigg( \frac{P_i P_j }{\Delta } - \hbar^2 \Delta \del_r \ln \bar{f}_i \del_r \ln f_j - B_i (\theta ) B_j (\theta ) - \hbar^2  \del_r \ln \bar{g}_i \del_r \ln g_j  \bigg) \\
&=&\! -\frac{1}{\rho^2 } \left( \frac{I_{ij} }{2} - B_i (\theta ) B_j (\theta ) - \frac{d\bar{S}_{\theta i }}{d\theta } \frac{dS_{\theta j }}{d\theta } \right) . 
\end{eqnarray}
The intensity of a collision $E^2 $ of $\Psi_1 $ and $\Psi_2 $ is 
\begin{eqnarray}
E^2 \!&=&\! - \frac{1}{\rho^2 } \Bigg\{ \bigg( \frac{I_{11} }{2} - B_1 (\theta ) ^2 - \left| \frac{dS_{\theta 1} }{d\theta } \right| ^2 \bigg) \left| \Psi_1 \right| ^2 + \bigg( \frac{I_{22} }{2} - B_2 (\theta ) ^2 - \left| \frac{dS_{\theta 2} }{d\theta } \right| ^2 \bigg) \left| \Psi_2 \right| ^2 \nonumber \\
	&& \ \ \ \ \ \  + \, {\rm Re}\bigg[ \bigg( I_{12} - 2 B_1 (\theta ) B_2 (\theta ) - 2 \frac{d\bar{S}_{\theta 1} }{d\theta } \frac{dS_{\theta 2} }{d\theta } \bigg) \bar{\Psi}_1 \Psi_2 \bigg] \Bigg\} \\
	&=&\! - \frac{1}{\rho^2 } \Bigg\{ \frac{I_{11} }{2} | \Psi_1 | ^2 + \frac{I_{22} }{2} | \Psi_2 | ^2 + {\rm Re}\Big( I_{12} \bar{\Psi }_1 \Psi_2 \Big) \nonumber \\
	&& \ \ \ \ \ \  - \bigg( B_1 (\theta ) ^2 + \left| \frac{dS_{\theta 1}}{d\theta} \right| ^2 \bigg) | \Psi_1 | ^2 - \bigg( B_2 (\theta ) ^2 + \left| \frac{dS_{\theta 2}}{d\theta} \right| ^2 \bigg) | \Psi_2 | ^2 \nonumber \\
	&& \ \ \ \ \ \  - 2 \, {\rm Re}\bigg[ \bigg( B_1 (\theta ) B_2 (\theta ) + \frac{d\bar{S}_{\theta 1} }{d\theta } \frac{dS_{\theta 2} }{d\theta } \bigg) \bar{\Psi }_1 \Psi_2 \bigg]  \Bigg\} . \label{E2arranged} 
\end{eqnarray}
Since $\Psi_1 $ and $\Psi_2 $ do not have physical divergences, the second and third lines of Eq.\,(\ref{E2arranged}) are nonsingular. Each of the first and second terms of the first line depends only on $\Psi_1 $ and $\Psi_2 $, respectively. Thus, these terms do not contain quantities relevant to the collision of the two field excitations. That is why our interest is the singular or nonsingular behavior of ${\rm Re}(I_{12} \bar{\Psi}_1 \Psi_2 )$. Using the arbitrariness of the argument of $C_0 ^{in} $ in $\Psi_2 $ (or $\Psi_1 $), i.e.,  
\begin{eqnarray}
C_0 ^{in} \rightarrow e^{i \alpha } C_0 ^{in} , 
\end{eqnarray}
we can get the imaginary part of $I_{12} \bar{\Psi}_1 \Psi_2 $ to be $0$: 
\begin{eqnarray}
{\rm Re}\Big( I_{12} \bar{\Psi }_1 \Psi_2 \Big) = | I_{12} | \, | \Psi_1 | \, |\Psi_2 | . 
\end{eqnarray}
Below, we calculate $I_{12} $ and discuss whether $| I_{12} | \, | \Psi_1 | \, |\Psi_2 |$ has singularities or not.

\subsection{Nonextremal case}\label{secIntensityNonext}
Let us calculate $I_{12} $ by local solutions of the Klein-Gordon equation in $r_H \ne r_C $. 
For an outgoing mode, from Eq.\,(\ref{fononext}), $\del_r \ln f^{out}$ is written as 
\begin{eqnarray}
\del_r \ln f^{out} \!\!\!&=&\!\!\! \frac{\beta }{2(r - r_H)} + \frac{\gamma }{2(r - r_C)} - \frac{\alpha }{2(r_H - r_C) } - \frac{\del_x \ln Hl(\alpha , \beta , \gamma , \delta , \eta ; x )}{r_H - r_C} \\
	\!\!\!&=&\!\!\! \frac{\beta }{2(r - r_H )} - \frac{2\eta + \beta}{2(r_H - r_C) (\beta + 1)} + {\mathcal O}(r - r_H) . 
\end{eqnarray}
\ \\
For an ingoing mode, $\del_r  \ln f^{in}$ is obtained through replacing $\beta$ by $- \beta$: 
\begin{eqnarray}
\del_r \ln f^{in} =  \frac{-\beta }{2(r - r_H )} - \frac{2\eta - \beta}{2(r_H - r_C) (-\beta + 1)} + {\mathcal O}(r - r_H) .  
\end{eqnarray}
\\
(i)When both modes are ingoing, 
\begin{eqnarray}
I_{12} \!&=&\! \frac{2}{\Delta } \bigg[ P_1 P_2 - \bigg( \frac{r - r_C}{r_H - r_C} \bigg)^2 P_{H1} P_{H2} \bigg] \nonumber \\
&&\ \ \ \ \ \  - \frac{\hbar^2 (2 \eta_1 - \bar{\beta}_1 ) \beta_2 }{2(1 - \bar{\beta}_1 )} - \frac{\hbar^2 (2\eta_2 - \beta_2 ) \bar{\beta}_1 }{2(1 - \beta_2 )} + \mathcal{O}(r - r_H ) . 
\end{eqnarray}
The first term of the right-hand side is an indeterminate form in the near-horizon limit $r \rightarrow r_H$ and has finite limit 
\begin{eqnarray}
&&\frac{2}{\Delta} \bigg[ P_1 P_2 - \bigg( \frac{r - r_C}{r_H - r_C} \bigg)^2 P_{H1} P_{H2} \bigg] \nonumber \\
	&&\ \ \  = \frac{-4P_{H1} P_{H2} }{(r_H - r_C)^2 } + \frac{2P^{\prime }_{H1} P_{H2} + 2P^{\prime }_{H2} P_{H1} }{r_H - r_C } + {\mathcal O}(r - r_H) \\
	&&\ \ \  = - \hbar^2 \bar{\beta}_1 \beta_2 - i \hbar P^{\prime }_{H1} \beta_2 + i \hbar P^{\prime }_{H2} \bar{\beta}_1 + \mathcal{O}(r - r_H ) . 
\end{eqnarray}
Therefore, we obtain 
\begin{eqnarray}
I_{12\,H} \!&=&\! \lim_{r \rightarrow r_H } I_{12} \\
&=&\! (r_H ^2 \mu_1 ^2 + \lam_1 - i \hbar P^{\prime } _{H1} ) \frac{\beta_2 }{1 - \bar{\beta_1 }} + (r_H ^2 \mu_2 ^2 + \lam_2 + i \hbar P^{\prime } _{H2} ) \frac{\bar{\beta}_1 }{1 - \beta_2 } .  \label{I12Hnonext}
\end{eqnarray}
Define
\begin{eqnarray}
X_i = \eps_i - \Omega_H m_i - \Phi_H e_i , 
\end{eqnarray}
so that 
\begin{eqnarray}
I_{12\,H} = (r_H ^2 \mu_1 ^2 + \lam_1 - i \hbar P^{\prime } _{H1} ) \frac{X_2 }{X_1 - i \hbar \kappa_H } + (r_H ^2 \mu_2 ^2 + \lam_2 + i \hbar P^{\prime } _{H2} ) \frac{X_1 }{X_2 + i \hbar \kappa_H } . \label{I12Hnonext3}
\end{eqnarray}
When $X_i \sim 0$ and $X_j \gg X_i $, then $|I_{12\,H} |$ takes a nearly maximum value: 
\begin{eqnarray}
|I_{12\,H} | \sim (r_H ^2 \mu_i ^2 + \lam_i - i \hbar P^{\prime } _{Hi} ) \frac{X_j }{\hbar \kappa_H } . 
\end{eqnarray}
Although it has been shown that in a nonextremal case the intensity of the collision $E^2 $ is bounded, the upper bound of $|I_{12\,H} |$ is proportional to $\displaystyle\frac{{\rm max}(X_1 , X_2 )}{\hbar \kappa_H }$ and large when $\hbar \kappa_H \ll {\rm max}(X_1 , X_2 )$. 

We can write down the real part and imaginary part explicitly as 
\begin{eqnarray}
\mathrm{Re}(I_{12\,H} ) \!\!&=&\!\! \left( r_H ^2 \mu_1 ^2 + \lam_1  \right) \frac{X_1 X_2 }{X_1 ^2 + (\hbar \kappa_H )^2 } + \hbar P^{\prime }_{H1} \frac{\hbar \kappa_H X_2 }{X_1 ^2 + (\hbar \kappa_H )^2} \nonumber \\
&& + \left( r_H ^2 \mu_2 ^2 + \lam_2  \right) \frac{X_1 X_2 }{X_2 ^2 + (\hbar \kappa_H )^2 } + \hbar P^{\prime }_{H2} \frac{\hbar \kappa_H X_1 }{X_2 ^2 + (\hbar \kappa_H )^2} , \label{noexReIHinin} \\
\mathrm{Im}(I_{12\,H} ) \!\!&=&\!\! \left( r_H ^2 \mu_1 ^2 + \lam_1  \right) \frac{\hbar \kappa_H  X_2 }{X_1 ^2 + (\hbar \kappa_H )^2 } - \hbar P^{\prime }_{H1} \frac{X_1 X_2 }{X_1 ^2 + (\hbar \kappa_H )^2} \nonumber \\
&& - \left( r_H ^2 \mu_2 ^2 + \lam_2  \right) \frac{\hbar \kappa_H X_1 }{X_2 ^2 + (\hbar \kappa_H )^2 } + \hbar P^{\prime }_{H2} \frac{X_1 X_2 }{X_2 ^2 + (\hbar \kappa_H )^2} . \label{noexImIHinin}
\end{eqnarray}

From Eqs.\,(\ref{defbeta}) and (\ref{defc}), 
\begin{eqnarray}
c = 1 - \beta . 
\end{eqnarray}
Equation (\ref{I12Hnonext}) is  
\begin{eqnarray}
I_{12\,H} = (r_H ^2 \mu_1 ^2 + \lam_1 - i\hbar P^{\prime } _{H1} ) \frac{1 - c_2 }{\bar{c}_1 } + (r_H ^2 \mu_2 ^2 + \lam_2 - i\hbar P^{\prime } _{H2} ) \frac{1 - \bar{c}_1 }{c_2 } . \label{I12Hnonext2}
\end{eqnarray}
When $\eps^2 = \mu^2 $ and $\eps (M \eps - Q e) = 0$, using $\bar{\G(z)} = \G(\bar{z})$, $c \G(c) = \G(c + 1)$, Eqs.\,(\ref{decomp}), (\ref{Nf1}) and (\ref{I12Hnonext2}), 
\begin{eqnarray}
\lim_{r \rightarrow r_H } \big( I_{12} \bar{\Psi }_1 \Psi_2 \big) \!\!\!&=&\!\!\! \left( \bar{C}_0 ^{in} \right)_1 \left( C_0 ^{in} \right)_2 (- x_0 )^{- \bar{A}_1 - A_2 } \bar{g}_1 (\theta ) g_2 (\theta ) e^{- i (\eps_2 - \eps_1 ) t + i (m_2 - m_1 ) \phi } \\
	&\times & \!\!\!\!\!\! \Bigg[ (r_H ^2 \mu_1 ^2 + \lam_1 - i\hbar P^{\prime } _{H1} ) \frac{\G(\bar{c}_1 - \bar{b}_1 ) \G(\bar{a}_1 )}{\G(\bar{c}_1 + 1) \G(\bar{a}_1 - \bar{b}_1 )} (1 - c_2 ) \frac{\G(c_2 - b_2 ) \G(a_2 )}{\G(c_2 ) \G(a_2 - b_2 )} \nonumber \\
	&&\!\!\!\!\!\!\! + \, (r_H ^2 \mu_2 ^2 + \lam_2 - i\hbar P^{\prime } _{H2} ) (1 - \bar{c}_1 ) \frac{\G(\bar{c}_1 - \bar{b}_1 ) \G(\bar{a}_1 )}{\G(\bar{c}_1 ) \G(\bar{a}_1 - \bar{b}_1 )} \frac{\G(c_2 - b_2 ) \G(a_2 )}{\G(c_2 + 1) \G(a_2 - b_2 )} \Bigg] . \nonumber  
\end{eqnarray}

\ \\
(ii)When mode $1$ is ingoing and mode $2$ is outgoing,
\begin{eqnarray}
I_{12} \!&=&\! \frac{2}{\Delta } \bigg[ P_1 P_2 + \bigg( \frac{r - r_C}{r - r_H} \bigg)^2 P_{H1} P_{H2} \bigg] - \frac{1}{2} \bigg( \frac{\beta_1 \beta_2 (2\eta_2 - 1)}{1 - \beta_2 ^2} + \frac{\beta_1 \beta_2 (2 \eta_1 - 1)}{1 - \beta_1 ^2} \bigg) \nonumber \\
 && - \frac{1}{2} \bigg( \frac{\beta_1 (\beta_2 ^2 - 2\eta_2)}{1 - \beta_2 ^2} + \frac{\beta_2 (\beta_1 ^2 - 2 \eta_1)}{1 - \beta_1 ^2} \bigg) + {\mathcal O}(r - r_H) \\
\!&\approx&\! \frac{4P_{H1} P_{H2}}{\Delta } + \mathcal{O}(\Delta^0 ) . \label{asyIinout}
\end{eqnarray}
It is shown that the stress tensor of a superposition of ingoing and outgoing modes diverges at the horizon, and comparing Eqs.\,(\ref{EcmNH2}) and (\ref{asyIinout}), it is found that near the horizon, the leading term of $g^{ab} \del_a \bar{S_1} \del_b S_2$ of ingoing and outgoing modes has no field-theoretical corrections.

\subsection{Extremal case}\label{secIntensityext}
From Eqs.\,(\ref{foext}) and (\ref{fiext}), the asymptotic behavior of $\del_r \ln f^{out}$ and $\del_r \ln f^{in}$ is written as 
\begin{eqnarray}
\del_{r} \ln f^{in} \!&\sim&\! - \frac{\alpha_{-1} }{2\xi^2 } + \frac{\beta_{-1} }{\xi } + \frac{\alpha_1 }{2} + \del_{\xi } \ln \sum_{n=0}^{\infty} \psi_n ^+ \bigg( \frac{\xi}{\alpha_{-1}} \bigg)^{n}  \\
\!&=&\! - \frac{\alpha_{-1} }{2\xi^2 } + \frac{\beta_{-1} }{\xi } + \frac{\alpha_1 }{2} + \frac{\psi^+ _1 }{\alpha_{-1}} + \mathcal{O}(\xi ) , \label{expandfin} \\ 
\del_{r} \ln f^{out} \!&\sim&\! \frac{\alpha_{-1} }{2\xi^2 } - \frac{\beta_{-1} }{\xi } + \frac{\alpha_1 }{2} + \del_{\xi } \ln \sum_{n=0}^{\infty} \psi_n ^- \bigg( \frac{\xi}{- \alpha_{-1}} \bigg)^{n} \ \\ 
\!&=&\! \frac{\alpha_{-1} }{2\xi^2 } - \frac{\beta_{-1} }{\xi } + \frac{\alpha_1 }{2} - \frac{\psi^- _1 }{\alpha_{-1}} + \mathcal{O}(\xi ) , 
\end{eqnarray}
where $\psi^{\pm}_1 $ have been obtained from Eq.\,(\ref{psipm}). 

\ \\
(i)When both modes are ingoing, 
\begin{eqnarray}
I_{12} \!&=&\! \frac{2 P_1 P_2 }{\xi^2 } - 2\hbar^2 \xi^2 \del_{r} \ln \bar{f}_1 ^{in} \del_{r} \ln f_2 ^{in} . \label{exI} 
\end{eqnarray}
The first term of the right-hand side of Eq.\,(\ref{exI}) is expanded as 
\begin{eqnarray}
\frac{P_1 P_2 }{\xi^2 } \!&=&\! \frac{P_{H1} P_{H2} }{\xi^2 } + \frac{P^{\prime }_{H1} P_{H2} + P^{\prime }_{H2} P_{H1} }{\xi } \nonumber \\
 && + \frac{1}{2} \Big(P^{\prime \prime }_{H1} P_{H2} + P^{\prime \prime }_{H2} P_{H1} \Big) + P^{\prime }_{H1} P_{H2} ^{\prime } + \mathcal{O}(\xi ) . \label{expandPP}
\end{eqnarray}
Using Eqs.\,(\ref{expandfin}) and (\ref{expandPP}), we obtain the near-horizon limit of $I_{12} $ : 
\begin{eqnarray}
I_{12\,H} = (M^2 \mu_1 ^2 + \lam_1) \frac{P_{H2}}{P_{H1}} + (M^2 \mu_2 ^2 + \lam_2) \frac{P_{H1}}{P_{H2}} + i \hbar \bigg( - \frac{P^{\prime }_{H1} P_{H2}}{P_{H1}} + \frac{P^{\prime }_{H2} P_{H1}}{P_{H2}} \bigg) . \label{exIHinin}
\end{eqnarray}

It can be checked that the result (\ref{exIHinin}) is consistent with the calculation in the nonextremal case Eq.\,(\ref{I12Hnonext3}). Equation (\ref{exIHinin}) indicates that the BSW effect occurs in the extremal limit. However, for the critical modes, $\alpha_{-1} = 0$, the asymptotic expansion Eqs.\,(\ref{foext}) and (\ref{fiext}) are ill defined. Fortunately, we obtained the solutions in such a situation in section \ref{secextcrit}. We will observe what happens for the critical modes in section \ref{tracecrit}. 

\ \\
(ii)When mode $1$ is ingoing and mode $2$ is outgoing, the leading term of the asymptotic behavior of $I_{12} $ to the horizon is the same as in the nonextemal case: 
\begin{eqnarray}
I_{12} \approx \frac{4P_{H1} P_{H2}}{\Delta } + o(\Delta ) . 
\end{eqnarray}

\subsubsection{Critical modes}\label{tracecrit}
For critical modes, from Eqs.\,(\ref{logderivf1crit}) and (\ref{logderivf2crit}), 
\begin{eqnarray}
\del_r \ln f_1 (\xi ) \!&=&\! \frac{s_+ }{\xi} + \kappa -2\kappa \frac{\alpha_+ }{\gamma_+ } \frac{F(\alpha_+ + 1 , \gamma_+ + 1 ; -2\kappa \xi)}{F(\alpha_+ , \gamma_+ ; -2\kappa \xi)} \\
  \!&=&\! \frac{s_+ }{\xi} + \kappa - 2\kappa \frac{\alpha_+ }{\gamma_+ } + \mathcal{O}(\xi ) , \\
\del_r \ln f_2 (\xi ) \!&=&\! \frac{s_- }{\xi} + \kappa - 2\kappa \frac{\alpha_- }{\gamma_- } + \mathcal{O}(\xi ) . 
\end{eqnarray}

\ \\
(i)When one mode is $f_2 (\xi )$ (say mode 1) and the other is ingoing but not critical(say mode 2), 
\begin{eqnarray}
I_{12} \!&=&\! - 2 \xi^2 \bigg( \frac{\bar{s}_- }{\xi} + \bar{\kappa } - 2\bar{\kappa } \frac{\bar{\alpha}_- }{\bar{\gamma}_- } \bigg)_1 \bigg( - \frac{\alpha_{-1} }{2\xi^2 } + \frac{\beta_{-1} }{\xi } + \frac{\alpha_1 }{2} + \frac{\psi_1 }{\alpha_{-1}} \bigg)_2 + \mathcal{O}(\xi ) \\
\!&\approx&\! \frac{ (\bar{s}_- )_1 (\alpha_{-1} )_2 }{\xi } + o(\xi^{-1}) . 
\end{eqnarray}
This situation is the BSW process for $P_{H1} = + 0$ or $(b_0 )_1 \le \frac{1}{4}$. 

As we saw in Table \ref{zeros}, when $b_0 \ge 0$, the amplitude at the horizon is not $0$. When $b_0 < 0$, the amplitude behaves as $\sim \xi^{\frac{-1 + \sqrt{1 - 4b_0 }}{2}}$. Therefore, if $b_0 > -2$ and $b_0 \ne 0$, then the intensity of the collision $E^2 \sim I_{12} \Psi_1 \Psi_2 $ diverges at the horizon. If $b_0 = 0$, then $s_- = 0$ so that $I_{12} < \infty $. 

In summary, if $b_0 > 0$ or $-2 < b_0 < 0$, then $I_{12} \times \Psi_1 $ diverges at the horizon.

\ \\
(ii)When one mode is $f_1 (\xi )$ (mode 1) and the other (mode 2) is ingoing but not critical, 
\begin{eqnarray}
I_{12} \approx  \frac{ (\bar{s}_ + )_1 ( \alpha_{-1})_2 }{\xi } + o(\xi^{-1}) . 
\end{eqnarray}
This situation is the BSW process for $P_H = - 0$ and $b_0 > \frac{1}{4}$. From Eq.\,(\ref{foextcrit}), 
\begin{eqnarray}
Nf_1 (\xi ) \!&\sim&\! C_0 ^{in} (-2\kappa )^{\gamma_+ - \alpha_+ } \frac{\Gamma(\alpha_+ )}{\Gamma(\gamma_+ )} \xi_0 ^{-s_+ - \alpha_+ + \gamma_+ } \xi^{s_+ } \\
I_{12} \Psi_1 \!&\sim&\! \xi^{s_+ - 1} \,=\, \xi^{\frac{-3 + i \sqrt{4b_0 - 1}}{2}} .  
\end{eqnarray}
Since $b_0 > \frac{1}{4}$, $I_{12} \Psi_1 $ diverges at the horizon. 

\ \\
(iii)When one mode is $f_2 (\xi )$ (mode 1) and the other (mode 2) is outgoing but not critical, 
\begin{eqnarray}
I_{12} \approx - \frac{(\bar{s}_- )_1 (\alpha_{-1})_2 }{\xi } + o(\xi^{-1}) . 
\end{eqnarray}

\ \\
(iv)When one mode is $f_1 (\xi )$ (mode 1) and the other (mode 2) is outgoing but not critical, 
\begin{eqnarray}
I_{12} \approx - \frac{(\bar{s}_+ )_1 (\alpha_{-1})_2 }{\xi } + o(\xi^{-1}) . 
\end{eqnarray}

\ \\
(v)When both of the modes are critical, 
\begin{eqnarray}
I_{12} = -2 \bar{s}_{\pm 1} s_{\pm 2} + \mathcal{O}(\xi ) . 
\end{eqnarray}
The quantity $I_{12}$ of the collision of critical modes is finite.


\section{Conclusion}\label{conclusion}
We have investigated the radial part of the Klein-Gordon equation for a massive and electrically charged scalar field in Kerr-Newman spacetime. In several specific cases, it reduces to an essentially hypergeometric equation, so that its exact solutions and their global nature are obtained in each case. The results of the reductions are summarized in Table \ref{reductions}. 
\begin{table}[!h]
\caption{Reductions of the radial part of the Klein-Gordon equation.}
\label{reductions}
\centering
\begin{tabular}{l|l|l}
\hline 
\multicolumn{1}{c}{Geometry} & \multicolumn{1}{|c|}{Conditions} & \multicolumn{1}{c}{Reduces to} \\
\hline \hline
$r_H \ne r_C $& No conditions & Confluent Heun equations \\
\hline
$r_H = r_C $& No conditions & Double confluent Heun equations \\
\hline 
$r_H \ne r_C $& $\eps^2 = \mu^2 $, \ \ \ \ \ \ \ \ \ \ \, $\eps (M \eps - Q e) = 0$ & Hypergeometric equations \\
\hline
 &$\eps^2 = \mu^2 = e Q = 0$& \\
$r_H \ne r_C $&$\eps^2 = \mu^2 = 0$, $MQe = - a m$& General Legendre equations \\
 &$\eps^2 = \mu^2 $, $M\eps = Q e$, $am = (M^2 - Q^2 )\eps $& \\
\hline 
$r_H = r_C $&$\ \ \ \ \ \ \ \ \ \ \ \, P_H = 0 $ &Confluent hypergeometric equations\\
\hline
$r_H = r_C $&$\eps^2 = \mu^2$, \ \ \ \ \ \ \ \ \ \ \ $\eps (M \eps - Q e) = 0$ &Confluent hypergeometric equations\\
\hline
$r_H = r_C $&$\eps^2 = \mu^2 $, $P_H = 0$&Bessel equations \\
\hline
$r_H = r_C $&$\eps^2 = \mu^2 $, $P_H = 0$, $\eps (M \eps - Q e) = 0$ &Euler equations \\
\hline
\end{tabular}
\end{table}

Using these global solutions and local solutions, we have evaluated the intensity $E^2 $ of ``collision'' of two field excitations. We expect that the intensity probes the back-reactions to the gravitational field. It has been shown that in a nonextremal case, $E^2$ of two ingoing modes is bounded. Comparing Eq.\,(\ref{EcmNH1}) with Eq.\,(\ref{I12Hnonext3}), the divergences in the classical particle calculation are regularized by the replacement 
\begin{eqnarray}
\frac{P_{Hi} }{P_{Hj} } = \frac{X_i }{X_j } \, \rightarrow \, \frac{X_i }{X_j \pm i \hbar \kappa_H } = \frac{X_i }{X_j } \left[ 1 \pm i \left( \displaystyle \frac{\hbar \kappa_H }{X_j } \right) \right] ^{-1} . \label{replacement}
\end{eqnarray}
In the limit $\hbar \kappa_H \rightarrow 0$, this term grows unboundedly if one mode is critical, namely $X_j = 0$. 

If the two modes are ingoing and outgoing, the intensity of the collision diverges at the event horizon no matter whether the black hole is extremal or not.  

Our calculations are restricted to the mode analysis in fixed stationary background geometries, i.e., Kerr-Newman geometries. To reveal what these unboundedly large quantities mean physically, further studies are necessary. In general, divergences of the stress tensor indicate that the fixed background approximation becomes invalid. Thus, from our results it is implied that under the existence of outgoing modes in any Kerr-Newman geometries, or critical modes in extremal geometries at the event horizon, the approximate prescription in which the background geometries are fixed ought to fail to describe the scalar fields near the horizon. 

In conclusion, our results suggest that the BSW effect is inherited by the quantum theory. However, the mechanism that restricts the unboundedly high-energy phenomena for ingoing modes within the extremal cases is quite different from that of classical particles. In classical particle theory, the potential barrier of a nonextremal black hole prevents critical particles from reaching the event horizon. On the other hand, in quantum theory, although critical modes reach the event horizon through the tunneling process, the intensities of the collisions are regularized by field-theoretical effects.

\section*{Acknowledgments}
The author would like to thank H. Kawai for fruitful discussions and some important comments.

\appendix

\section{Relations between the solutions}
In this section, we see the relations between the solutions we obtained in sections \ref{Nonextmarginal}, \ref{gLegendre}, \ref{secextcrit}, \ref{secmarginal}, \ref{seccritmarginal}, and \ref{seccritmarginalspec}. 

To avoid confusion, we write $f^{out} (\xi )_{\rm (marginal)} $ and $f^{in} (\xi )_{\rm (marginal)}$ for the solutions (\ref{fopure}) and (\ref{fipure}) that we obtained in section \ref{secmarginal} and $f_1 (\xi )_{\rm (critical)} $ and $f_2 (\xi )_{\rm (critical)} $ for Eqs.\,(\ref{foutPH=0}) and (\ref{finPH=0}) in section \ref{secextcrit}. 

\subsection{Nonextremal case}
Let us see the relation between sections \ref{Nonextmarginal} and \ref{gLegendre}. 

First, in case (i) $\eps^2 = \mu^2 = Q e = 0$, 
\begin{eqnarray}
a \!&=&\! - i (2M\eps - Q e) + \frac{1}{2} \left\{ 1 - i \sqrt{ 4\big[ 2\eps P_H + (2M\eps - Q e)^2 - r_H ^2 \mu^2 - \lam \big] - 1 } \right\} \\
&=&\! \frac{1}{2}  \Big[ 1 - i \sqrt{- 4\lam - 1}  \Big] \\
&=&\! 1 + \nu , 
\end{eqnarray}
Similarly, 
\begin{eqnarray}
b \!&=&\! \frac{1}{2} \Big[ 1 + i \sqrt{- 4\lam - 1}  \Big] \\
&=&\! - \nu , \\
c \!&=&\! 1 - n 
\end{eqnarray}
and 
\begin{eqnarray}
\Lambda_1 = - \Lambda_2 = - \frac{n}{2} . 
\end{eqnarray}
Therefore, from Eq.\,(\ref{deff1}), 
\begin{eqnarray}
f_1 \!&=&\! \left( \frac{1 + u}{1 - u} \right)^{\frac{n}{2} } F\big(1 + \nu , - \nu ; 1 - n ; \textstyle \frac{1 - u}{2} \big) \\
	&=&\! \G(1 - n)\, f_P ^{+} (u) . 
\end{eqnarray}
Using the linear transformation formula~\cite{HMF1964}
\begin{eqnarray}
F(a , b ; c ; z) = (1 - z)^{c - a - b} F(c - a , c - b ; c ; z) , \label{LTF1}
\end{eqnarray}
Eq.\,(\ref{deff2}) is 
\begin{eqnarray}
f_2 \!&=&\! x^{- \Lambda_1 } (1 - x)^{\Lambda_2 + c - a - b} F(1 - a , 1 - b ; 2 - c ; x) \\
	&=&\! \left( \frac{1 + u}{1 - u} \right)^{- \frac{n}{2}} F\big( - \nu , 1 + \nu ; 1 + n ; \textstyle \frac{1 - u}{2} \big) \\
	&=&\! \G(1 + n)\, f_P ^{-} (u) .   
\end{eqnarray}

\ \\

Next, in both case (ii) and (iii), namely, Eqs.\,(\ref{marginal2b}) and (\ref{marginal2c}) respectively, $P_H = - P_C $ and 
\begin{eqnarray}
a \!&=&\! - \frac{ i (P_H - P_C )}{r_H - r_C } + \frac{1}{2} \left\{ 1 - i \sqrt{ 4\big[ 2\eps P_H + (2M\eps - Q e)^2 - r_H ^2 \mu^2 - \lam \big] - 1 } \right\} \\
	&=&\! - \frac{2 i P_H }{r_H - r_C } + \frac{1}{2} \left\{ 1 - i \sqrt{ 4\big[ 2\eps P_H + (2M\eps - Q e)^2 - r_H ^2 \mu^2 - \lam \big] - 1 } \right\} \\
	&=&\! \left\{ 
	\begin{array}{l}
	- n + \frac{1}{2} \Big[ 1 - i \sqrt{4 Q^2 e^2 - 4\lam - 1}  \Big] \ \ \ \ \ \ \ \ \ \ \ \ \ \ \ \ \ \ \ \ \ \ \ \ \ \, {\rm ( \,case\ (ii) \, ) , } \\
	- n + \frac{1}{2}  \Big[ 1 - i \sqrt{- 4 (M^2 - a^2 - Q^2 ) \eps^2 - 4\lam - 1}  \Big] \ \ \ \ \ \ {\rm ( \, case\ (iii) \, ) } 
	\end{array}
	\right. \\
	&=&\! - n + 1 + \nu . 
\end{eqnarray}
Similarly, 
\begin{eqnarray}
b \!&=&\! - n - \nu , \\
c \!&=&\! 1 - n , \\
\Lambda_1 \!\!&=&\!\! \Lambda_2 \, =\, - \frac{n}{2} . 
\end{eqnarray}
From Eqs.\,(\ref{deff1}), (\ref{deff2}), and formula (\ref{LTF1}), we obtain 
\begin{eqnarray}
f_1 \!&=&\! \left( \frac{1 - u}{2} \right)^{- \frac{n}{2}} \left( \frac{1 + u}{2} \right)^{- \frac{n}{2}} F\big( -n + 1 + \nu , - n - \nu ; 1 - n ; \textstyle \frac{1 - u}{2} \big) \\
 	&=&\! \left( \frac{1 + u}{1 - u} \right)^{\frac{n}{2}} F\big( - \nu , 1 + \nu ; 1 - n ; \textstyle \frac{1 - u}{2} \big) \\
	&=&\! \G(1 - n)\, f_P ^+ (u)  
\end{eqnarray}
and 
\begin{eqnarray}
f_2 \!&=&\! \left( \frac{1 + u}{1 - u} \right)^{- \frac{n}{2}} F\big( 1 + \nu , - \nu ; 1 + n ; \textstyle \frac{1 - u}{2} \big) \\ 
	&=&\! \G(1 + n)\, f_P ^- (u) .  
\end{eqnarray}

\subsection{The extremal limit of $f_1 $ and $f_2 $}
In sections \ref{Nonextmarginal} and \ref{secmarginal}, the conditions for the parameters of the field excitations Eqs.\,(\ref{marginal1}) and (\ref{extmarginal}) are equivalent. Therefore, it is expected that the extremal limits of $f_1 $ and $f_2 $ are $f^{in} (\xi )_{\rm (marginal)}$ and $f^{out} (\xi )_{\rm (marginal)}$, respectively, up to the scale factors. For a consistency check, let us calculate the extremal limit of Eqs.\,(\ref{nonextf1inf}) and (\ref{nonextf2inf}). 

From Eq.\,(\ref{nonextf1inf}), 
\begin{eqnarray}
&& (-1)^{\frac{i P_H }{r_H - r_C }} (r_H - r_C )^{-i (2M\eps - Q e)} f_1 \\
&=&\!\! (- 1)^{\frac{i P_H }{r_H - r_C }} (r_H - r_C )^{-i (2M\eps - Q e )} x^{\Lambda_1 } (1 - x)^{\Lambda_2 } \nonumber \\
&&\!\! \times \bigg[ (r_H - r_C)^a \frac{\G(c)}{\G(c - a)} \frac{\G(b - a)}{\G(b)} (r - r_H )^{- a} F\big( a, 1 - c + a; 1 - b + a; x^{-1} \big) \label{nonextf1infb} \\
&& + \, (r_H - r_C)^b \frac{\G(c)}{\G(c - b)} \frac{\G(a - b)}{\G(a)} (r - r_H )^{- b} F\big( b, 1 - c + b; 1 - a + b; x^{-1} \big)  \bigg] . \nonumber   
\end{eqnarray}
The extremal limit of the first line of (\ref{nonextf1infb}) can be calculated as 
\begin{eqnarray}
&& (- 1)^{\frac{i P_H }{r_H - r_C }} (r_H - r_C )^{-i (2M\eps - Q e )} x^{\Lambda_1 } (1 - x)^{\Lambda_2 }  \\
	&=&\!\!\! ( r - M )^{- \frac{i (P_H - P_C )}{r_H - r_C }} \Big( 1 + \frac{M - r_H }{r - M} \Big)^{\frac{r - M}{M - r_H } \frac{i P_H }{2(r - M)}} \Big( 1 + \frac{M - r_C }{r - M} \Big)^{\frac{r - M}{M - r_C } \frac{i P_C }{2(r - M)}}  \\
&\longrightarrow &\!\!\! \left( r - M \right)^{- i (2M\eps - Q e)} \exp\left( \frac{i P_H }{r - M } \right) \\
&=&\! \xi^{-i(2M\eps - Q e ) } e^{\frac{\kappa }{\xi }} ,  
\end{eqnarray}
where ``$\longrightarrow $'' means the extremal limit. The limit of $a$ is 
\begin{eqnarray}
a \!\!\!&=&\!\!\! - i (2M\eps - Q e) + \frac{1}{2} \left\{ 1 - i \sqrt{4\big[ 2\eps P_H + (2M\eps - Q e)^2 - r_H ^2 \mu^2 - \lam \big] - 1 } \right\} \\
	&\longrightarrow &\! -i P^{\prime }_H + \frac{1}{2} \left\{ 1 - i \sqrt{4 [P^{\prime \prime }_H P_H + (P^{\prime }_H )^2 - M^2 \mu^2 -\lam ] -1 } \right\} \\
	&=& \alpha_- . 
\end{eqnarray}
Similarly, in the same limit, 
\begin{eqnarray}
&&\!\!\!\! b \longrightarrow \alpha_+ , \\
&&\!\!\!\! 1 + a - b \longrightarrow \gamma_- , \\
&&\!\!\!\! 1 + b - a \longrightarrow \gamma_+ , \\
&&\!\!\!\! i(2M\eps - Q e) + a \longrightarrow s_- , \\
&&\!\!\!\! i(2M\eps - Q e) + b \longrightarrow s_+ . 
\end{eqnarray}
The hypergeometric functions $F(a, b; c; z)$ are related to the confluent hypergeometric functions $F(a, c; z)$ as 
\begin{eqnarray}
\lim_{b \rightarrow \infty } F\Big( a , b ; c ; \frac{z}{b} \Big) = \lim_{b \rightarrow \infty } F\Big( b , a ; c ; \frac{z}{b} \Big) = F(a , c ; z) . 
\end{eqnarray}
Thus, 
\begin{eqnarray}
F\big( a, 1 - c + a ; 1 - b + a ; x^{- 1} \big) \longrightarrow F\Big( \alpha_- , \gamma_- ; - \frac{2\kappa }{\xi }  \Big) , \\
F\big( b, 1 - c + b ; 1 - a + b ; x^{- 1} \big) \longrightarrow F\Big( \alpha_+ , \gamma_+ ; - \frac{2\kappa }{\xi }  \Big) . \label{HGandCHG}
\end{eqnarray}
From the asymptotic expansion of the gamma function~\cite{HMF1964}, 
\begin{eqnarray}
z^{\beta - \alpha } \frac{\G(z + \alpha )}{\G(z + \beta )} \sim 1 + \frac{(\alpha - \beta ) (\alpha + \beta - 1)}{2 z} + \mathcal{O}(z^{-2} ) . \label{asymptG/G}
\end{eqnarray}
Using this formula, 
\begin{eqnarray}
(r_H - r_C )^a \frac{\G(c)}{\G(c - a)} \!&=&\! (r_H - r_C )^a \frac{\G(c -1 + 1)}{\G(c - 1 + 1 - a)} \\
	&=&\! (r_H - r_C )^a (c - 1)^a (c - 1)^{- a} \frac{\G(c - 1 + 1)}{\G(c - 1 + 1 - a)} \\
	&=&\! (- 2 i P_H )^a (c - 1)^{- a} \frac{\G(c - 1 + 1)}{\G(c - 1 + 1 - a)} \\
	&\longrightarrow & (-2 \kappa )^{\alpha_- } , 
\end{eqnarray}
and 
\begin{eqnarray}
(r_H - r_C )^b \frac{\G(c)}{\G(c - b)} \longrightarrow (-2 \kappa )^{\alpha_+ } . 
\end{eqnarray}
Therefore, we obtain the extremal limit of (\ref{nonextf1infb}): 
\begin{eqnarray}
(-1)^{\frac{i P_H }{r_H - r_C }} (r_H - r_C )^{-i (2M\eps - Q e) } f_1 \longrightarrow f^{in} (\xi ) _{\rm (marginal)} , 
\end{eqnarray}
where $f^{in} (\xi ) _{\rm (marginal)} $ is Eq.\,(\ref{fipure}). 

\ \\

By similar calculations, one can derive the extremal limit of $f_2 $. From Eq.\,(\ref{nonextf2inf}), 
\begin{eqnarray}
&& (- 1)^{- \frac{i P_H }{r_H - r_C }} (r_H - r_C )^{i(2M\eps - Q e)} f_2 \\
	&=&\!\! (- 1)^{- \frac{i P_H }{r_H - r_C }} (r_H - r_C )^{i(2M\eps - Q e)} x^{- \Lambda_1 } (1 - x)^{\Lambda_2 } (- x)^{c - 1} \nonumber \\
	&&\!\! \times \bigg[ (r_H - r_C )^a \frac{\G(2 - c)}{\G(1 + b - c)} (r - r_H )^{- a} F\big( 1 - c + a, a; 1 + a - b; x^{- 1} \big) \\
	&& + (r_H - r_C )^b \frac{\G(2 - c)}{\G(1 - c + a)} (r - r_H )^{- b} F\big( 1 + b - c, b; 1 - a + b; x^{- 1} \big) \bigg] . \nonumber 
\end{eqnarray}
Using the relation (\ref{HGandCHG}) and the asymptotic formula (\ref{asymptG/G}), one finds 
\begin{eqnarray}
(- 1)^{- \frac{i P_H }{r_H - r_C }} (r_H - r_C )^{i (2 M \eps - Q e )} f_2 \longrightarrow f^{out} (\xi ) _{\rm (marginal)} ,  
\end{eqnarray}
where $f^{out} (\xi )_{\rm (marginal)} $ is Eq.\,(\ref{fopure}).

\subsection{Extremal case}
The relation between the confluent hypergeometric function and the Bessel function is~\cite{HMF1964} 
\begin{eqnarray}
\lim_{a \rightarrow \infty } \left[ \frac{F(a , b ; - \frac{z}{a})}{\G(b)} \right] = z^{\frac{1}{2} - \frac{b}{2}} J_{b - 1} \big( 2 \sqrt{z} \big) . 
\end{eqnarray}
Using this relation, the marginal limit $\kappa \rightarrow 0$ of Eq.\,(\ref{finPH=0}) is  
\begin{eqnarray}
\lim_{\kappa \rightarrow 0} f _2 (\xi ) _{\rm (critical)} \!&=&\! \lim_{\kappa \rightarrow 0} \xi^{s_- } e^{\kappa \xi } F( \alpha_- , \gamma_- ; - 2 \kappa \xi ) \\
	&=&\! \lim_{\kappa \rightarrow 0} \xi^{s_- } e^{\kappa \xi } F\Big( \frac{b_1 }{2\kappa } , \gamma_- ; - \frac{2 \kappa }{b_1 } b_1 \xi \Big) \\
	&=&\! (b_1 )^{\frac{i}{2} \sqrt{4b_0 - 1} } \G(\gamma_- ) \xi^{- \frac{1}{2}} J_{-i \sqrt{4b_0 - 1}} \big( 2 \sqrt{b_1 \xi } \big) \\
	&=&\! (b_1 )^{\frac{i}{2} \sqrt{4b_0 - 1} } \G(\gamma_- ) \xi^{- \frac{1}{2}} \tilde{f}_- (2 \sqrt{b_1 \xi }) ,  
\end{eqnarray}
where $J_{-i \sqrt{4b_0 - 1}} \big( 2 \sqrt{b_1 \xi } \big)$ and $\tilde{f}_- (2 \sqrt{b_1 \xi })$ are defined by Eq.\,(\ref{defBessel}). Using the asymptotic behavior of $\tilde{f}_- (y)$ Eq.\,(\ref{asympextp0Bessel}), 
\begin{eqnarray}
\lim_{b_1 \rightarrow 0} \lim_{\kappa \rightarrow 0} f_2 (\xi ) _{\rm (critical)} = \xi^{n_- } , 
\end{eqnarray}
where $n_{\pm } = - \frac{1}{2} \pm \frac{i}{2} \sqrt{4b_0 - 1}$. 

Similarly, 
\begin{eqnarray}
\lim_{\kappa \rightarrow 0} f_1 (\xi ) _{\rm (critical)} = (b_1 )^{- \frac{i}{2} \sqrt{4b_0 - 1} } \G(\gamma_+ ) \xi^{- \frac{1}{2}} \tilde{f}_+ (2 \sqrt{b_1 \xi })  
\end{eqnarray}
and
\begin{eqnarray}
\lim_{b_1 \rightarrow 0} \lim_{\kappa \rightarrow 0} f_1 (\xi ) _{\rm (critical)} = \xi^{n_+ } . 
\end{eqnarray}

\bibliographystyle{h-elsevier}
\bibliography{sample}

\end{document}